\documentclass[fleqn,usenatbib]{mnras}

\usepackage{newtxtext,newtxmath}

\usepackage[T1]{fontenc}

\DeclareRobustCommand{\VAN}[3]{#2}
\let\VANthebibliography\thebibliography
\def\thebibliography{\DeclareRobustCommand{\VAN}[3]{##3}\VANthebibliography}

\usepackage{graphicx}	
\usepackage{amsmath}	

\usepackage{pifont}

\newcounter{relation}
\renewcommand{\therelation}{\alph{relation}}

\newcommand{\relation}{%
  \refstepcounter{relation}%
  \therelation\label{rel:\therelation}%
}

\newcommand{\relationbf}{%
  \refstepcounter{relation}%
  \textbf{\therelation}\label{rel:\therelation}%
}

\newcommand{\orcid}[1]{%
  \href{https://orcid.org/#1}{%
    \raisebox{0.3ex}{\hspace{0.12em}\includegraphics[height=7pt]{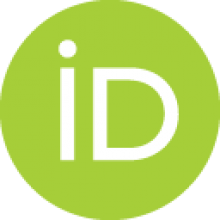}}%
  }%
}

\title[Binary black hole merger times from cosmology]{Scaling Relations for Binary Black Hole Merger Times from Cosmological Initial Conditions}

\author[T. L. Fran\c{c}ois et al.]{
Thibaut L. Fran\c{c}ois\orcid{0009-0001-0314-7038},$^{1}$\thanks{E-mail: thibaut.francois@proton.me}
Alessia Gualandris\orcid{0000-0002-9420-2679}$^{1}$
and Walter Dehnen\orcid{0000-0001-8669-2316}$^{2}$
\\
$^{1}$School of Mathematics and Physics, University of Surrey, Guildford, GU2 7XH, UK\\
$^{2}$Astronomisches Rechen-Institut, Zentrum für Astronomie der Universität Heidelberg, Mönchhofstraße 12-14, 69120, Heidelberg, Germany\\
}

\date{Accepted 2026 July 7. Received 2026 May 18.}

\pubyear{\the\year{}}

\begin{document}
\label{firstpage}
\pagerange{\pageref{firstpage}--\pageref{lastpage}}
\maketitle

\begin{abstract}
Recent evidence from Pulsar Timing Arrays (PTAs) for a nanohertz gravitational wave background is broadly consistent with theoretical expectations from a population of massive black hole binaries (MBHBs), although the inferred amplitude appears somewhat higher than predicted by standard models. Interpreting these observations requires a robust understanding of the merger timescales of MBHBs, and of their connection to host galaxy properties.
In this work, we investigate the evolution of MBHBs selected from cosmological galaxy mergers in the \textsc{IllustrisTNG} simulation. We re-simulate these systems at high resolution using the $N$-body code \textsc{Griffin} to accurately resolve the dynamical friction and stellar hardening phases, and follow their evolution to coalescence with a semi-analytical model.
We find that cosmological galaxy encounters and the resulting MBHBs are typically highly eccentric. We characterise the distribution of binary eccentricities at formation and at entry into the PTA band, and quantify the corresponding residence times.
We identify the key parameters governing the duration of the different stages of MBHB evolution, and derive scaling relations linking galaxy and orbital properties to dynamical friction, hardening, and total coalescence times. These relations provide a framework for subgrid prescriptions in cosmological simulations.
Applying these scaling relations to the full \textsc{IllustrisTNG} merger population, we infer the probability distributions of galaxy merger and black hole coalescence times. We find that galaxy mergers typically complete within $\sim 0.7$ Gyr, while the total black hole coalescence time is $\sim 1.0$ Gyr. These short timescales imply efficient binary evolution, consistent with current PTA constraints.
\end{abstract}




\section{Introduction}
\label{sec:intro}

Supermassive black holes (SMBHs), with masses ranging from $\sim10^6$ to $10^{10} M_\odot$, are widely believed to reside at the centres of most massive galaxies \citep[][]{Kormendy1995, Ferrarese2005, Kormendy2013}. In the $\Lambda$ cold dark matter paradigm, galaxies assemble hierarchically through a sequence of mergers and accretion events \citep[][]{Ostriker1975, Ostriker1977, WhiteRees1978, Lacey1993}. As a consequence, galaxy mergers naturally lead to interactions between their central SMBHs and to the formation of supermassive black hole binaries \citep[SMBHBs;][]{Begelman1980}.

The orbital evolution of an SMBH pair formed during a galaxy merger is commonly described as a three-phase process, governed by different mechanisms that extract energy and angular momentum from the system. In the first phase, dynamical friction against the surrounding dark matter and stellar background efficiently drives the SMBHs from kiloparsec scale separations down to parsec scales \citep[][]{Chandrasekhar1943}. In the second phase, further orbital decay is driven by repeated three-body interactions between the binary and individual stars in the nuclear region, which are ejected through the gravitational slingshot mechanism, leading to binary hardening \citep[][]{Hills1983, Quinlan1996}. Finally, at sub-parsec separations, gravitational wave (GW) emission becomes the dominant mechanism driving the binary evolution, rapidly bringing the SMBHs to coalescence \citep[][]{Peters1963, Peters1964}. The efficiency and duration of each phase depend sensitively on both the properties of the host galaxies and the orbital parameters of the merger, making SMBHB evolution a complex and dynamical process.

For binaries with chirp masses $> 10^8M_\odot$ (see equation~\ref{eq:chirpMass}), the inspiral phase emits GWs predominantly at nanohertz frequencies, giving rise to a stochastic gravitational wave background (GWB) generated by the incoherent superposition of a large population of unresolved SMBHBs throughout cosmic history \citep[][]{Sesana2008}. Pulsar timing arrays (PTAs) are uniquely sensitive to this low-frequency GW regime and have recently reported evidence for a nanohertz GWB \citep[][]{Agazie2023a, EPTA2023, Reardon2023, Xu2023}. The observed signal is broadly consistent with a background dominated by inspiralling SMBHBs, but the measured amplitude sits at the high end of current theoretical expectations based on standard assumptions for SMBH demographics and merger histories. If the observed GWB is primarily produced by SMBHBs, this discrepancy may point to a population that is more massive, more numerous, or evolves more efficiently toward coalescence than typically assumed \citep[][]{Agazie2023a, Agazie2023b, EPTA2024}. 

One source of this tension lies in the way the GWB is commonly estimated. Current predictions typically combine large cosmological simulations with semi-analytic models, in which the dynamics of SMBHs is not resolved explicitly. In these simulations, black holes are often repositioned at the centres of their host potentials and assumed to undergo prompt mergers, effectively equating the SMBH merger rate to the galaxy merger rate \citep[][]{Schaye2015, Weinberger2018}. This approach neglects the detailed binary evolution and the delays associated with SMBHB formation and hardening, though we note that not all cosmological simulations adopt this simplification. More elaborate approaches introduce delays associated with dynamical friction acting on the infalling SMBH \citep[see][]{Tremmel2015, Barausse2020, Ma2023, QuelquejayLeclere2026}. However, the subsequent evolution of the bound SMBH binary is still not resolved. Instead, its hardening timescale is typically estimated using analytic prescriptions calibrated on idealised $N$-body simulations and expressed in terms of host galaxy properties at the binary sphere of influence \citep[e.g.][]{Sesana2015}. These recipes are widely adopted despite the fact that the sphere of influence is far below the resolution limit, requiring galaxy properties to be extrapolated from much larger scales.

As a consequence, predictions of the GWB are affected by substantial uncertainties, as the amplitude and spectral shape of the background depend sensitively on the merger timescales of SMBHBs \citep[][]{Agazie2023b}. These timescales are, in turn, set by the detailed dynamical evolution of the binary and therefore by the properties of the host galaxies and the orbital parameters of their encounter. A robust understanding of how SMBHB merger timescales depend on galaxy structure and merger dynamics is thus essential for improving semi-analytic prescriptions and, ultimately, for producing more reliable predictions of the GWB.

Establishing a quantitative link between host galaxy properties and SMBHB merger timescales is, however, intrinsically challenging, as it requires large suites of high resolution $N$-body simulations in order to capture the relevant dynamical processes and build statistically meaningful samples, at a substantial computational cost. As a step in this direction, \citet{HolleyBockelmann2025} derived empirical relations between SMBHB merger timescales and host galaxy properties based on a set of 28 idealised $N$-body simulations, corresponding to 9 distinct galaxies. While pioneering, their approach adopts several simplifying assumptions that facilitate the numerical modelling but may limit its applicability to realistic cosmological mergers. Specifically, the setup follows the inspiral of a single SMBH into a host galaxy rather than modelling two self-consistently merging galaxies, neglects dark matter and gas, and assumes a fixed initial orbital eccentricity of $e=0.5$, rather than the range expected in cosmological encounters.\citep[see][]{Khochfar2006, Fastidio2024, Gualandris2026}.

In this paper, we aim to complement these efforts and establish a connection between the properties of galaxy mergers and the coalescence timescales of SMBH binaries, using realistic initial conditions drawn directly from the \textsc{IllustrisTNG} simulation. We identify the key parameters that govern the different stages of the binary evolution and construct scaling relations linking the timescales of the various evolutionary phases to both host galaxy properties and the orbital parameters of the merger. The remainder of this paper is organised as follows:

\noindent $\blacksquare$ \, \hyperref[sec:methods]{\textbf{\textcolor{black}{Section 2:}}} We describe our methodology, including the selection of galaxy mergers from \textsc{IllustrisTNG}, the extraction of initial conditions, the setup of our $N$-body simulations, and the semi-analytical model used to evolve SMBH binaries up to coalescence.\\
\noindent $\blacksquare$ \, \hyperref[sec:times]{\textbf{\textcolor{black}{Section 3:}}} We define the characteristic timescales that describe the key stages of binary evolution.\\
\noindent $\blacksquare$ \, \hyperref[sec:results]{\textbf{\textcolor{black}{Section 4:}}} We present our results, starting with the statistics of cosmological mergers and the properties of the simulated binaries, including the time spent in each dynamical phase. 
We then focus on the detectability of these binaries by PTA experiments. We subsequently describe our procedure for deriving scaling relations and present the resulting relations. Finally, we apply these relations to the full \textsc{IllustrisTNG} merger sample to derive the coalescence time distribution of supermassive black hole binaries.\\
\noindent $\blacksquare$ \, \hyperref[sec:conclusion]{\textbf{\textcolor{black}{Section 5:}}} We summarise the main results of the paper, compare them with previous work, discuss key caveats, and provide a practical guide for applying our scaling relations.\\

\section{Methods}
\label{sec:methods}

\subsection{Initial conditions from cosmological simulations}
\label{sec:cosmoIC}

\subsubsection{\textsc{IllustrisTNG} simulations}
\label{sec:tngSims}

\textsc{IllustrisTNG} is a suite of large volume cosmological gravo-magnetohydrodynamical simulations performed with the moving-mesh code \textsc{AREPO} \citep[][]{Springel2010}. The simulation suite consists of three distinct volumes, TNG50, TNG100, and TNG300, corresponding to cubic boxes of $51.7^3$, $110.7^3$, and $302.6^3$ cMpc$^3$, respectively \citep[][]{Springel2018, Pillepich2018b, Naiman2018, Nelson2018, Marinacci2018, Nelson2019a, Nelson2019, Pillepich2019}.

Each TNG simulation self-consistently follows the coupled evolution of dark matter, gas, stars, and SMBHs from an initial redshift of $z=127$ down to the present day, $z=0$, storing the evolution in 100 snapshots. All runs start from cosmologically motivated initial conditions assuming a $\Lambda$CDM cosmology consistent with the Planck 2015 results \citep[][]{Planck2015}. 

Supermassive black holes in the TNG simulations are initially seeded with a mass of $8\times 10^5\, h^{-1}M_\odot$\footnote{Here, $h$ denotes the dimensionless Hubble parameter, with 
$h=0.6774$.} in dark matter haloes exceeding $5\times 10^{10}\, h^{-1}M_\odot$ \citep[][]{Pillepich2018a}. Once seeded, the black holes grow via gas accretion following a Bondi–Hoyle prescription, with the accretion rate limited by the Eddington threshold. SMBHs remain anchored to the minimum of their host galaxy's potential and undergo prompt binary mergers \citep[][]{Weinberger2017, Weinberger2018}.

In order to build a statistically significant sample of galaxy and black hole mergers, we select merger events from all three TNG volumes. For each volume, we use the highest resolution realizations, TNG50-1, TNG100-1, and TNG300-1, which contain $2\times 2160^3$, $2\times 1820^3$ and $2\times 2500^3$ resolution elements, respectively. For dark matter and stellar particles, the softening length is fixed in comoving units at high redshift and transitions to a constant physical value below $z = 1$. The comoving softening is set to $\epsilon_{\rm com} = L_{\rm box}/N_{\rm DM}^{1/3}/40$, corresponding to a physical softening, $\epsilon_{\rm phys}(z) = \frac{\epsilon_{\rm com}}{1+z}$, which is frozen to its $z=1$ value at lower redshifts. 

For the highest resolution realisations TNG50-1, TNG100-1, and TNG300-1, this prescription corresponds to $0.29$, $0.74$, and $1.48$ kpc, respectively, at redshifts $z<1$. While the mass and spatial resolution of TNG300-1 are lower than those of TNG50-1 and TNG100-1, its inclusion is necessary to access a sufficiently large number of merger events and to construct a statistically meaningful sample spanning a broad range of galaxy and encounter parameters. At this resolution, global quantities such as total masses, mass ratios, and orbital parameters are captured and remain consistent with the cosmology but the inner density structure of galaxies is less well constrained.

For a detailed description of the numerical implementation, physical models, parameter choices, and validation of the \textsc{IllustrisTNG} simulations, we refer the reader to the TNG methods papers by \citet{Weinberger2017} and \citet{Pillepich2018a}, as well as to the official \textsc{IllustrisTNG} documentation (\url{https://www.tng-project.org/}).

\subsubsection{Merger selection in \textsc{IllustrisTNG}}
\label{sec:mergerSel}

We select galaxies whose properties are consistent with the expected hosts of PTA sources, namely massive systems harbouring SMBHs in the mass range accessible to current and future PTA experiments. We first select, at redshift $z=0$, all galaxies with a stellar mass $M_\star > 3\times 10^{11} M_\odot$, yielding 33, 149, and 1552 systems in TNG50, TNG100, and TNG300, respectively. We then use the \textsc{Sublink} merger trees \citep[][]{RodriguezGomez2015}, which track the assembly history of structures at the subhalo (i.e. galactic) level, to reconstruct the merger histories of these galaxies. From this sample, we identify all major mergers, defined by a stellar mass ratio $q_\star > 1/4$, occurring at redshifts $z \leq 1$. This redshift and mass ratio selection is motivated by the expectation that the stochastic GWB is dominated by low-redshift major merger events \citep[][]{Sesana2008, IzquierdoVillalba2022, Agazie2023b}. For each selected encounter, we compute the orbital parameters of the interacting galaxies (see Section~\ref{sec:extractIC} for details) and discard mergers associated with unbound orbits. This results in 256 galaxy mergers across the three simulation volumes.

For each merger event, the properties of the progenitor galaxies are generally extracted from the snapshot immediately preceding the merger. In some cases, however, one of the progenitors is not identified in this snapshot, typically because its density contrast with respect to the surrounding background becomes too low. When this occurs, we trace both progenitor branches further back in time until a snapshot is found in which the two galaxies are simultaneously identified. Merger events for which no such common snapshot exists are excluded from the final sample. During this procedure, the mass ratio may change slightly, as the progenitor properties are evaluated at snapshots located further back in time. For 13 merger events in our initial selection, the revised mass ratio no longer satisfies the condition $q_\star > 1/4$, and these systems are therefore removed from the sample.

We then verify that both galaxies host a SMBH, and that the corresponding SMBHB satisfies a mass ratio $q_{\rm BH} > 1/4$ and a chirp mass ${\cal M} > 10^{8}\ M_\odot$ \citep[][]{Sesana2009, IzquierdoVillalba2022, Agazie2023b}, where 

\begin{equation}
    \mathcal{M} = \frac{(M_1\, M_2)^{3/5}}{(M_1+M_2)^{1/5}},
	\label{eq:chirpMass}
\end{equation}
with $M_1$ and $M_2$ the masses of the two black holes.
This results in 74 relevant mergers. Given our limited computational resources and the fact that our $N$-body simulation code does not include gas dynamics (see Section~\ref{sec:simCode}), we apply a final selection based on the gas content. We retain only mergers for which the gas mass within twice the stellar half-mass radius is less than 30 per cent of the stellar mass, yielding a final sample of 30 mergers for our study.

\subsubsection{Extraction of the cosmological initial conditions}
\label{sec:extractIC}

\begin{figure*}
	\centering
	\includegraphics[width=\textwidth]{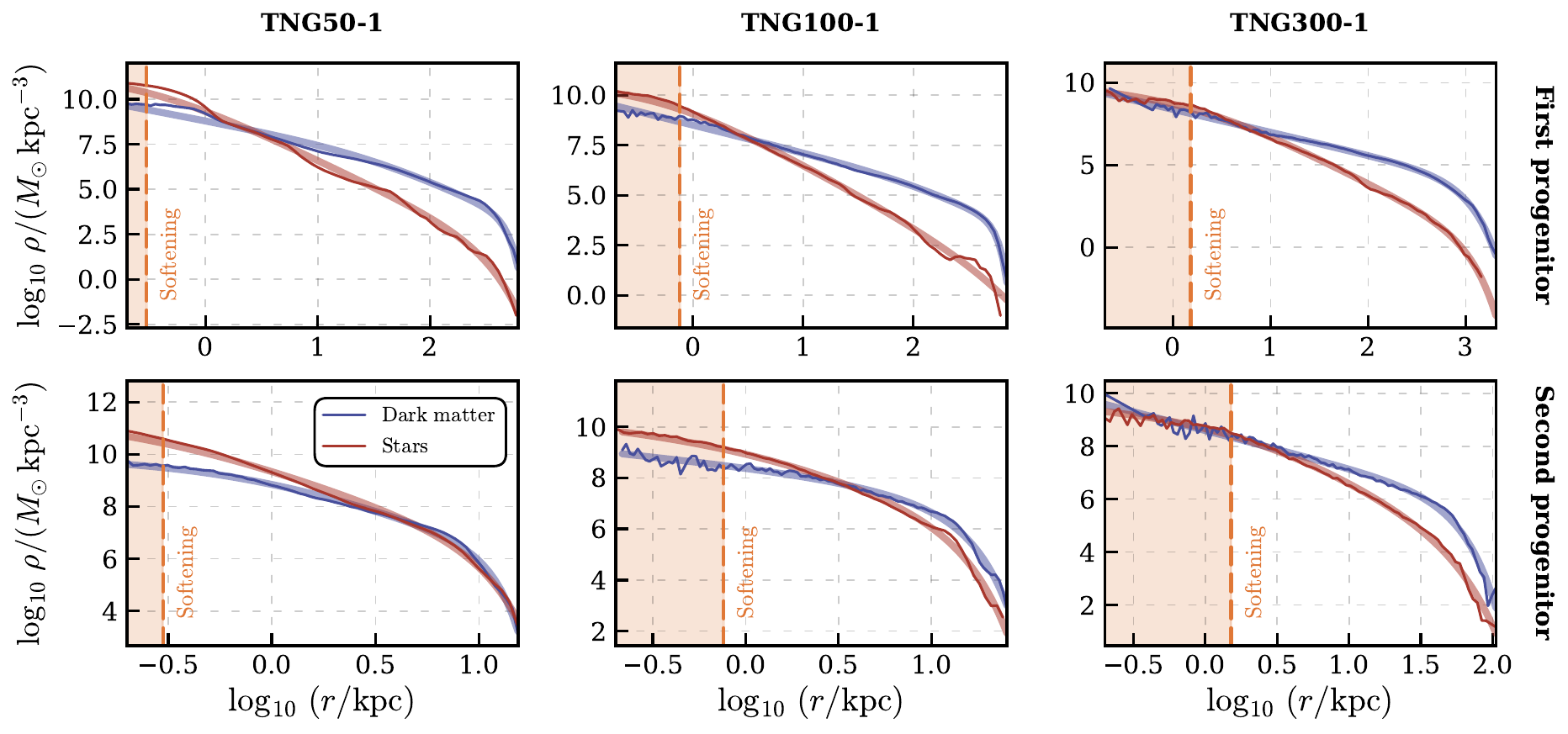}
    \caption{Density profile fits for galaxies extracted from the \textsc{IllustrisTNG} simulations. The original TNG density profiles are shown as thin, dark lines, while the corresponding best-fitting truncated spheroidal models are displayed as thick, lighter lines. Dark matter and stellar components are shown in blue and red, respectively. Each column presents an example pair of galaxies drawn from one of the three \textsc{IllustrisTNG} volumes. The top and bottom rows correspond to the first and second progenitors, respectively.}
    \label{fig:densityTNG}
\end{figure*}

For each merger, we fit the stellar and dark matter density profiles of the progenitors using a truncated spheroidal model of the form:

\begin{equation}
    \rho(r) = \rho_0 \, \left(\frac{r}{r_\mathrm{s}}\right)^{-\gamma} \left[ 1 + \left(\frac{r}{r_\mathrm{s}}\right)^\alpha \right]^{(\gamma-\beta)/\alpha} \exp\!\left[-\left(\frac{r}{r_\mathrm{cut}}\right)^\xi \right]
    \label{eq:denProfile}
\end{equation}
where $(r_\mathrm{s}, \rho_0)$ are the scale radius and the corresponding density at that radius, respectively. We set the outer slope $\beta=3$ and the transition slope $\alpha=1$ while the inner slope $\gamma$ is left as a free parameter. The parameters $(r_{\rm cut}, \xi)$ control the truncation radius and the steepness of the exponential cutoff, respectively. The truncation allows the outer regions of the profile to be adjusted in order to account for matter stripping induced by the cosmological environment from which the initial conditions are extracted. The fit therefore has five free parameters: $\rho_0$, $r_{\rm s}$, $\gamma$, $r_{\rm cut}$ and $\xi$, all of which are constrained to be positive. Examples of the stellar and dark matter density profile fits are shown in Fig.~\ref{fig:densityTNG}.

To characterise the orbital configuration of each encounter, we approximate the interacting galaxies as a Keplerian two-body system. Each galaxy is treated as a point mass located at its centre, moving with its bulk velocity, such that the relative separation $r_{\mathrm{rel}}$ and velocity $v_{\mathrm{rel}}$ fully describe the orbit. Under this assumption, the semi-major axis $a_0$ and the orbital eccentricity $e_0$ of the encounter are computed as
\begin{equation}
a_0 = \left( \frac{2}{r_{\mathrm{rel}}} - \frac{v_{\mathrm{rel}}^{2}}{G M} \right)^{-1} \quad ; \quad e_0 = \sqrt{1 - \frac{h^{2}}{G M a_0}}
\label{eq:galacticOrbit}
\end{equation}
where $M$ denotes the total enclosed mass of the two galaxies within the relative separation, $G$ is the gravitational constant and $h$ is the angular momentum per unit mass. The pericentre $r_{\mathrm{peri}}$ and apocentre $r_{\mathrm{apo}}$ are then given by
\begin{equation}
r_{\mathrm{peri}} = a_0 (1 - e_0) \quad ; \quad r_{\mathrm{apo}} = a_0 (1 + e_0).
\end{equation}

\subsection{$N$-body simulations}
\label{sec:Nbody}

\subsubsection{Sampling from \textsc{IllustrisTNG} profiles}
\label{sec:samplingIC}

Initial conditions for the stellar and dark matter components are generated using the \textsc{Agama} library \citep[][]{Vasiliev2019}. Each galaxy is represented by an equilibrium distribution function constructed self-consistently from the density profiles and gravitational potentials obtained from fits to the \textsc{IllustrisTNG} simulations. The distribution functions are action-based and assume a quasi-spherical geometry. Phase-space samples are drawn directly from these distribution functions. This procedure allows us to resample the TNG galaxies at higher resolution.

\subsubsection{Choice of the number of particles}
\label{sec:partNb}

The choice of the total number of particles is non-trivial, particularly with respect to the behaviour of the black hole binary eccentricity at formation, $e_\mathrm{b}$, which is known to exhibit significant scatter. Recent work has suggested that this scatter may partly reflect a genuine physical sensitivity to small perturbations in the nearly radial plunging trajectories of the SMBHs prior to binary formation \citep{Rawlings2023}. However, numerical resolution also plays an important role: \citet{Gualandris2026} have shown that simulations with insufficient resolution can suffer from additional stochasticity in $e_\mathrm{b}$, driven by fluctuating torques from the stellar distribution during pericentric passages. The required resolution to control this stochasticity increases with the initial orbital eccentricity of the galaxy encounter, $e_0$, although the precise dependence is further modulated by additional parameters, including the galaxy masses, mass ratio, and central densities, which determine the number of particles within the binary's sphere of influence. As a result, predicting a priori the resolution needed to achieve a given level of convergence in $e_\mathrm{b}$ is challenging.

This issue is particularly relevant for our sample, which comprises 30 mergers spanning a wide range of initial conditions in terms of $e_0$, total mass, mass ratio, and internal density structure. While \citet{Gualandris2026} report an approximately one-to-one correspondence between $e_0$ and converged $e_\mathrm{b}$ for encounters with $0.9 < e_0 < 0.97$, this result was obtained for a specific galaxy model without dark matter, and its applicability outside this interval or to more complex galaxy models remains uncertain. In particular, the value of $e_0$ above which this relation breaks down is expected to depend on the adopted galactic model. We therefore use this relation primarily as a qualitative guideline rather than a strict convergence criterion.

We therefore determine the total number of particles, $N_\mathrm{tot}$, individually for each merger by balancing two competing requirements. First, we aim to maximise the number of particles of the first progenitor within five times the binary influence radius at $t=0$, in order to minimise stochastic effects on $e_\mathrm{b}$, particularly for highly eccentric encounters that penetrate deep into the central potential, even during the first pericentric passage. In practice, this corresponds to typically $(4-12)\times10^{5}$ particles within this region (see Section~\ref{sec:massRefine} and \ref{sec:Resolution} for details). Second, we require the total integration time to remain computationally feasible, as the mergers are initialised at apocentre and some systems start at very large separations, reaching distances of up to $479$ kpc. Further discussion of numerical resolution is provided in Section~\ref{sec:massRefine}.

Once the total number of particles has been determined, they are distributed such that the first progenitor contains equal numbers of stellar and dark matter particles. Combined with the mass refinement scheme, this practical choice enhances the resolution of the stellar component while ensuring that individual dark matter particles are sufficiently light to avoid introducing spurious perturbations to the binary. We further ensure that the particle masses are the same in both the first and second progenitors and that the mass ratio is preserved. Table~\ref{tab:particleNb} summarises the number of particles used for each run.

\begin{table}
    \centering
    \caption{Summary of the numerical resolution for each simulation run. We list the total number of particles ($N_\mathrm{tot}$), the number of dark matter particles ($N_\mathrm{DM}$), and the number of stellar particles ($N_\star$).}
    \label{tab:particleNb}
    \begin{tabular}{l c l c c c}
        \hline
        Merger & $N_\mathrm{tot}$ & Galaxy & $N_\star$ & $N_\mathrm{DM}$ \\
        \hline
        1 & $4 \times 10^6$ & Primary   & 1671925 & 1671925 \\
          & & Secondary & 645302 & 10852   \\
        2 & $4 \times 10^6$ & Primary   & 1717529 & 1717529 \\
          & & Secondary & 550399  & 14543   \\
        3 & $4 \times 10^6$ & Primary   & 1619350 & 1619350 \\
          & & Secondary & 661254  & 100049   \\
        4 & $6 \times 10^6$ & Primary   & 2496670 & 2496670 \\
          & & Secondary & 950795  & 55866   \\
        5 & $4 \times 10^6$ & Primary   & 1756545 & 1756545 \\
          & & Secondary & 480765  & 6144   \\
        6 & $7 \times 10^6$ & Primary   & 2986888 & 2986888 \\
          & & Secondary & 1006521  & 19705   \\
        7 & $5 \times 10^6$ & Primary   & 2154506 & 2154506 \\
          & & Secondary & 644130  & 46862   \\
        8 & $5 \times 10^6$ & Primary   & 2099932 & 2099932 \\
          & & Secondary & 781290  & 18849   \\
        9 & $4 \times 10^6$ & Primary   & 1717699 & 1717699 \\
          & & Secondary & 532865  & 31742   \\
        10 & $9 \times 10^6$ & Primary   & 3742320 & 3742320 \\
          & & Secondary & 1489598  & 25763   \\
        11 & $7 \times 10^6$ & Primary   & 3054325 & 3054325 \\
          & & Secondary & 875414  & 15938   \\
        12 & $7 \times 10^6$ & Primary   & 2438094 & 2438094 \\
          & & Secondary & 2041820  & 81993   \\
        13 & $9 \times 10^6$ & Primary   & 4077634 & 4077634 \\
          & & Secondary & 824731  & 20001   \\
        14 & $8 \times 10^6$ & Primary   & 3359305 & 3359305 \\
          & & Secondary & 1259989  & 21405   \\
        15 & $7 \times 10^6$ & Primary   & 3092618 & 3092618 \\
          & & Secondary & 802684  & 12081   \\
        16 & $12 \times 10^6$ & Primary   & 5188862 & 5188862 \\
          & & Secondary & 1604555  & 17724   \\
        17 & $4 \times 10^6$ & Primary   & 1433013 & 1433013 \\
          & & Secondary & 1111629  & 22349   \\
        18 & $4 \times 10^6$ & Primary   & 1634475 & 1634475 \\
          & & Secondary & 691638  & 39411   \\
        19 & $6 \times 10^6$ & Primary   & 2391115 & 2391115 \\
          & & Secondary & 1186189  & 31583   \\
        20 & $5 \times 10^6$ & Primary   & 2086524 & 2086524 \\
          & & Secondary & 778052  & 48900   \\
        21 & $6 \times 10^6$ & Primary   & 2464183 & 2464183 \\
          & & Secondary & 1044163  & 27474   \\
        22 & $5 \times 10^6$ & Primary   & 1914539 & 1914539 \\
          & & Secondary & 1144203  & 26723   \\
        23 & $6 \times 10^6$ & Primary   & 2419292 & 2419292 \\
          & & Secondary & 1088987  & 72431   \\
        24 & $4 \times 10^6$ & Primary   & 1657352 & 1657352 \\
          & & Secondary & 670924  & 14376   \\
        25 & $6 \times 10^6$ & Primary   & 1471002 & 1471002 \\
          & & Secondary & 998252  & 59748   \\
        26 & $4 \times 10^6$ & Primary   & 1707865 & 1707865 \\
          & & Secondary & 571327  & 12945   \\
        27 & $8 \times 10^6$ & Primary   & 3128138 & 3128138 \\
          & & Secondary & 1713616  & 30109   \\
        28 & $8 \times 10^6$ & Primary   & 3089130 & 3089130 \\
          & & Secondary & 1772566  & 49178   \\
        29 & $9 \times 10^6$ & Primary   & 3475657 & 3475657 \\
          & & Secondary & 2030801  & 17887   \\
        30 & $10 \times 10^6$ & Primary   & 3784663 & 3784663 \\
          & & Secondary & 2334255  & 96419   \\
          
        \hline
    \end{tabular}
\end{table}

\subsubsection{Mass refinement}
\label{sec:massRefine}

To enhance the mass resolution in the inner regions of the galaxies, we apply a mass refinement scheme to the particles of each progenitor. Rather than directly sampling the target number of particles, we initially generate a larger particle set and subsequently reassign particle masses in a controlled manner. Particles are first ranked according to their pericentric distance and divided into four pericentre shells. The innermost shell retains the highest mass resolution, while particles in progressively outer shells are downsampled and assigned larger masses such that the total mass, the original density profile and the target total number of particles $N_\mathrm{tot}$ are preserved.

This procedure results in eight distinct particle populations (four for the stellar component and four for the dark matter halo) with increasing particle mass as a function of pericentric distance. The mass refinement is governed by three parameters: (i) the number of shells, (ii) the fraction of the total particle population assigned to each shell, and (iii) the multiplicative factor by which particle masses increase from the innermost to the outermost shells. The specific values adopted in this work are summarised in Table~\ref{tab:massRefine}.

\begin{table}
    \centering
    \caption{Parameters defining the mass refinement scheme used in our simulations. We list for each shell the relative particle mass multiplier, and the fraction of the total particle population assigned.}
    \label{tab:massRefine}
    \begin{tabular}{c c c}
        \hline
        Shell index & mass multiplier & fraction of population \\
        \hline
        1 & 1  & 0.05 \\
        2 & 3  & 0.45 \\
        3 & 10 & 0.35 \\
        4 & 40 & 0.15 \\
        \hline
    \end{tabular}
\end{table}

Our primary goal is to achieve very high mass resolution in a compact central region, as several mergers in our sample are highly eccentric and involve close passages through the galaxy centres, where insufficient resolution would lead to spurious numerical effects. This requires balancing the need for maximal central resolution against the risk of introducing excessively massive particles in the outskirts, which could in principle perturb the inner dynamics. We tested several refinement configurations and converged on the parameters listed in Table~\ref{tab:massRefine} which provide excellent resolution in the central regions, with particle masses of order a few $10^{4}\,M_\odot$, while keeping the masses of outer-shell particles sufficiently low. Details on the particle number in the central region and on the mass of the light and heavy particles are provided in Section~\ref{sec:Resolution}. 

Long-term isolated tests spanning several gigayears show that massive outer particles never migrate into the central regions (details on the softening used in Section~\ref{sec:simCode}), which remain almost exclusively populated by particles from the innermost shell. In merger simulations, the second progenitor is initially placed well inside the innermost radius reached by the most massive particles of the first progenitor, ensuring that the SMBHs never interact with these particles. Although some massive particles from the second progenitor may plunge into the primary galaxy, they represent on average only $\sim 0.1\%$ of the total particle population, and only a small fraction of them ever approach the centre. Interactions with the SMBH binary are therefore extremely unlikely, particularly given that the binary hardens on timescales much shorter than the dynamical times of these outer particles. In total, we performed 107 simulations as part of this study and did not observe any interaction between the SMBH binary and massive outer-shell particles that could artificially affect its evolution.

\subsubsection{Orbital insertion and starting point}
\label{sec:orbitIC}

The two galaxies are placed on an initial orbit defined by the eccentricity and semi-major axis extracted from \textsc{IllustrisTNG}. All simulations are initialised at apocentre. This choice maximises the duration over which tidal interactions act during the first approach, allowing the galaxies to respond to the orbital evolution prior to the first pericentric passage.

Starting the simulations at apocentre also provides a well-defined and orbit-based reference point for measuring merger timescales. Since the orbital parameters are the primary quantities used to characterise and compare different encounters, initialising the systems at the same orbital phase ensures a consistent definition of the merger clock across the full sample and avoids ambiguities associated with arbitrary initial separations. This choice also provides additional flexibility, as the time origin can be redefined a posteriori to start all mergers at a common initial separation if desired.

\subsubsection{$N$-body code and softening}
\label{sec:simCode}

We model the dynamical evolution of the selected mergers using \textsc{Griffin}, a high-performance collisionless $N$-body code based on the fast multipole method \citep[FMM;][]{Dehnen2014}. The code is designed to maintain tight control over gravitational force errors through continuous monitoring and an adaptive choice of numerical parameters, resulting in a well-behaved error distribution comparable to that of direct summation techniques. Gravitational interactions among stars and dark matter particles are computed using the FMM, which retains $\mathcal{O}(N^{0.87})$ scaling, while interactions involving SMBHs are treated via direct summation in order to accurately capture their collisional dynamics and close encounters. 

As discussed in Section~\ref{sec:massRefine}, our simulations include multiple particle populations with different masses. To mitigate artificial mass segregation and the sinking of massive particles due to spurious collisional effects, we adopt a mass-dependent gravitational softening, in which more massive particles are assigned larger softening lengths.

For the stellar and dark matter components, the softening length is defined as\footnote{The scaling $\varepsilon \propto m^{1/2}$ is motivated by the requirement that the acceleration induced by close particle–particle encounters remains much smaller than that generated by the smooth mean-field potential, consistent with the collisionless nature of the system.}

\begin{equation}
    \varepsilon = \varepsilon_0 \times \left ( \frac{m}{m_0} \right)^{1/2}
\end{equation}
where $\varepsilon_0$ is a reference softening length, set to $\varepsilon_0=20$ pc, $m$ is the particle mass, and $m_0$ corresponds to the mass of the most centrally refined (i.e. lightest) dark matter particles. With this choice, stellar particles typically have softenings smaller than $\varepsilon_0$, ranging from $\sim 2-3$ pc up to $\sim 20$ pc, while dark matter particles have softenings spanning from $\varepsilon_0$ up to $\sim 125$ pc for the most massive particles.

For the SMBHs, we adopt a constant softening length $\varepsilon_\mathrm{BH,df}$ during the early phase of the simulation, dominated by dynamical friction. This value is then reduced to $\varepsilon_\mathrm{BH,bin}$ after the third pericentric passage, shortly before the formation of a bound binary. Specifically, we set $\varepsilon_\mathrm{BH,df} = 10\ \mathrm{pc}$ and $\varepsilon_\mathrm{BH,bin} = 2\ \mathrm{pc}$.

\subsection{Integration until coalescence}
\label{sec:coalescence}

To follow the late evolution of the SMBH binaries, we use a semi-analytic model to integrate the coupled differential equations governing the orbital elements under the combined effects of stellar hardening and gravitational wave (GW) emission. The time evolution of the semi-major axis $a$ and eccentricity $e$ can be written as the sum of the stellar and GW contributions,

\begin{equation}
\frac{da}{dt} = \left.\frac{da}{dt}\right|_\star + \left.\frac{da}{dt}\right|_\mathrm{GW} \quad \hbox{and} \quad \, \frac{de}{dt} = \left.\frac{de}{dt}\right|_\star + \left.\frac{de}{dt}\right|_\mathrm{GW}.
\label{eq:SAM}
\end{equation}
The first term (denoted $\star$) accounts for three-body interactions with surrounding stars, while the second (denoted GW) describes orbital decay driven by GW radiation. The stellar contribution will be extracted from our $N$-body simulations and is expressed as

\begin{equation}
\left.\frac{da}{dt}\right|_{\star} = -s(t)\, a^{2}  \quad \hbox{and} \quad \, \left.\frac{de}{dt}\right|_{\star} = s(t)Ka,
\label{eq:dade}
\end{equation}
where $s(t)$ and $K$ denote the time-dependent hardening rate and the eccentricity growth rate of the binary, respectively, defined as \citep[][]{Quinlan1996},

\begin{equation}
s(t) = \frac{d}{dt}\left(\frac{1}{a}\right)
\quad \hbox{and} \quad \, K = \frac{de}{d\ln(1/a)}.
\end{equation}
The contribution from GW emission is computed using Peters' equations \citep[][]{Peters1964}. This yields
\begin{equation}
\left.\frac{da}{dt}\right|_\mathrm{GW} = -\frac{64}{5} \frac{G^3 M_1 M_2 (M_1+M_2)} {c^5 a^3(1-e^2)^{7/2}} \left(1+\frac{73}{24}e^2+\frac{37}{96}e^4\right),
\label{eq:peters_a}
\end{equation}

\begin{equation}
\left.\frac{de}{dt}\right|_\mathrm{GW} = -\frac{304}{15}\, e\, \frac{G^3 M_1 M_2 (M_1+M_2)} {c^5 a^4(1-e^2)^{5/2}} \left(1+\frac{121}{304}e^2\right).
\label{eq:peters_e}
\end{equation}

In practice, for the stellar hardening contribution we obtain $s(t)$ and $K$ directly from our simulations. To estimate $s(t)$, we bin the time series of $1/a(t)$ and compute the local slope within each bin, which provides a discrete estimate of the hardening rate as a function of time. The resulting $s(t)$ measurements are then fitted with a decaying exponential function using an MCMC, with each bin weighted by the uncertainty on the corresponding slope. For the eccentricity growth, we find that the relation between $e$ and $\ln(1/a)$ is well described by a linear function. We therefore determine $K$ by performing a linear fit to $e(\ln(1/a))$ and extracting the corresponding slope.

\section{Binary evolutionary times}
\label{sec:times}

To characterise the key stages of the SMBH merger process, we define four characteristic times as follows. 

\subsection{Binding time}
\label{sec:tb}

We first introduce $t_\mathrm{b}$, which marks the end of the dynamical friction phase and the formation of a bound binary. This time is defined as the moment when the binding energy of the SMBH pair,
\begin{equation}
    E_\mathrm{b} = \frac{1}{2}\,\frac{M_1 M_2}{(M_1 + M_2)}
\,v_{\mathrm{rel}}^{2}
- \frac{G\,M_1 M_2}{d_{\mathrm{rel}}},
\end{equation}
becomes negative and remains so for the rest of the simulation. Here, $M_1$ and $M_2$ denote the masses of the two black holes, while $d_\mathrm{rel}$ and $v_\mathrm{rel}$ are their relative distance and relative velocity, respectively.

\subsection{Hard separation time}
\label{sec:th}

We define the \emph{hard separation time}\footnote{Not to be confused with the \emph{hardening time}, which refers to the total duration of the hardening phase.} as the time at which the binary becomes hard. This quantity serves as a useful diagnostic of the binary evolution (see Section~\ref{sec:tmerg}), but is not used to derive merger timescales. Following \citet{MerrittBook2013}, this time can be defined in two different ways, based either on the binary binding energy or on its semi-major axis.

The first definition, based on the binding energy, defines $t_\mathrm{hE}$ as the time at which the binding energy per unit mass of the binary equals the stellar velocity dispersion in the central region of the galaxy,
\begin{equation}
    \frac{E_\mathrm{b}(t=t_\mathrm{hE})}{(M_1+M_2)} = \sigma_\star^2 \left(<\frac{r_\mathrm{hm}}{2}\right)
\end{equation}
where $\sigma_\star$ denotes the stellar velocity dispersion measured within half the stellar half-mass radius $r_{\rm hm}$.

The second definition, based on the semi-major axis, defines $t_\mathrm{ha}$ as the time at which the binary separation satisfies
\begin{equation}
    d_\mathrm{rel}(t=t_\mathrm{ha}) = a_\mathrm{h} \quad \hbox{with} \quad a_\mathrm{h} = \frac{q_\mathrm{BH}}{(1+q_\mathrm{BH})^2} \frac{r_\mathrm{inf}}{4}
\end{equation}
where $r_\mathrm{inf}$ is the binary influence radius, defined as the radius enclosing a stellar mass equal to twice the mass of the more massive black hole in the primary progenitor \citep[][]{MerrittBook2013}, and $q_\mathrm{BH}$ is the black hole mass ratio.

The two definitions generally yield different values of the hard separation time, with the energy-based definition typically occurring at later times than the semi-major-axis-based one. We find that the latter provides a more practical and robust indicator in most cases. However, it is not always defined: in particular, mergers with highly eccentric initial orbits may coalesce before reaching $a_\mathrm{h}$.

\subsection{Gravitational wave dominated time}
\label{sec:tgw}

The time $t_\mathrm{gw}$ is defined as the moment when the rate of the change of the semi-major axis from stellar hardening and GW become equal, i.e. when
\begin{equation}
\left.\frac{da(t=t_{\rm gw})}{dt}\right|_\star = \left.\frac{da(t=t_{\rm gw})}{dt}\right|_\mathrm{GW}.
\end{equation}
It therefore marks the transition between the stellar hardening phase and the regime dominated by general relativistic effects.

\subsection{Coalescence time and uncertainties}
\label{sec:tmerg}

Equations~(\ref{eq:SAM}) are integrated using a fifth order Runge-Kutta scheme until one of the following conditions is met: (i) the binary pericentre falls below $10\, r_{\rm sch}$\footnote{We have verified that the choice of the prefactor in front of $r_{\rm sch}$ has no impact on the inferred coalescence time for any value up to a factor of $\sim 50$.}, with $r_{\rm sch}=2GM_{\rm bin}/c^2$ the Schwarzschild radius of the binary with total mass $M_{\rm bin}$, or (ii) the total integration time exceeds $50$~Gyr\footnote{Equations~(\ref{eq:peters_a}) and (\ref{eq:peters_e}) rely on an orbit-averaged formalism and are therefore formally valid only when the radiation-reaction timescale $t_{\rm rad}=a/|\dot{a}_{\rm GW}|$ exceeds the orbital period $T_{\rm orb}$. We have verified that this condition is satisfied throughout the integration for all mergers in our sample, with $t_{\rm rad}/T_{\rm orb} \gtrsim 100$, in all but two cases. The exceptions are mergers 5 and 20, which form with extremely high eccentricity and coalesce immediately after binary formation. For these systems, the orbit-averaged approximation becomes marginally violated near coalescence; however, given the extremely rapid evolution of the binary at this stage, the impact on the inferred coalescence time is expected to be negligible.}. If the integration terminates due to criteria (i), the corresponding time, $t_{\rm coal}$ is taken as the total coalescence time of the SMBH binary. Systems reaching the upper time limit (criterion ii) are instead classified as non merging.

Several methodological choices in the semi-analytic integration can affect the inferred coalescence times. 
First, to account for the uncertainties on the fit of $s(t)$ we draw 200 realisations from the posterior distributions of the fit parameters obtained via MCMC. Second, in rapidly evolving mergers the limited number of available data points can introduce a mild dependence of the inferred $s(t)$ on the adopted binning. To account for this effect, we repeat the estimation of $s(t)$ using five different choices for the number of bins. Finally, we explore different initial conditions for the integration. While an early starting point would ideally capture relativistic effects as soon as possible, the binary initially undergoes a transient phase during which its orbital parameters are not yet well settled. In practice, we find that $t_\mathrm{ha}$ provides a reasonable compromise\footnote{In the few cases where $t_\mathrm{ha}$ is not defined, we adopt $t_\mathrm{hE}$ instead.}. We therefore initialise the integrations from the first five data points following $t_\mathrm{ha}$.

This procedure results in a total of 5000 integrations per merger, yielding a distribution of coalescence times. We adopt the median of this distribution as the fiducial coalescence time, and use the 16th and 84th percentiles to quantify the associated uncertainty.

\subsection{Definition of timescales}
\label{sec:tmscalesDef}

We define the characteristic timescales associated with the different phases of SMBHB evolution as follows. The dynamical friction timescale, $\Delta t_{\rm df}$, corresponds to the time interval between the beginning of the simulation and the formation of a bound binary, i.e. $\Delta t_{\rm df} = t_{\rm b}$. The hardening timescale, $\Delta t_{\rm h}$, is defined as the time elapsed between binary formation and the onset of the GW-driven regime, such that $\Delta t_{\rm h} = t_{\rm gw} - t_{\rm b}$. Finally, the GW timescale, $\Delta t_{\rm gw}$, is defined as the remaining time until coalescence, i.e. $\Delta t_{\rm gw} = t_{\rm coal} - t_{\rm gw}$.

\section{Results}
\label{sec:results}

\subsection{Statistics of \textsc{IllustrisTNG} mergers}
\label{sec:tngStat}

\begin{figure*}
	\centering
	\includegraphics[width=\textwidth]{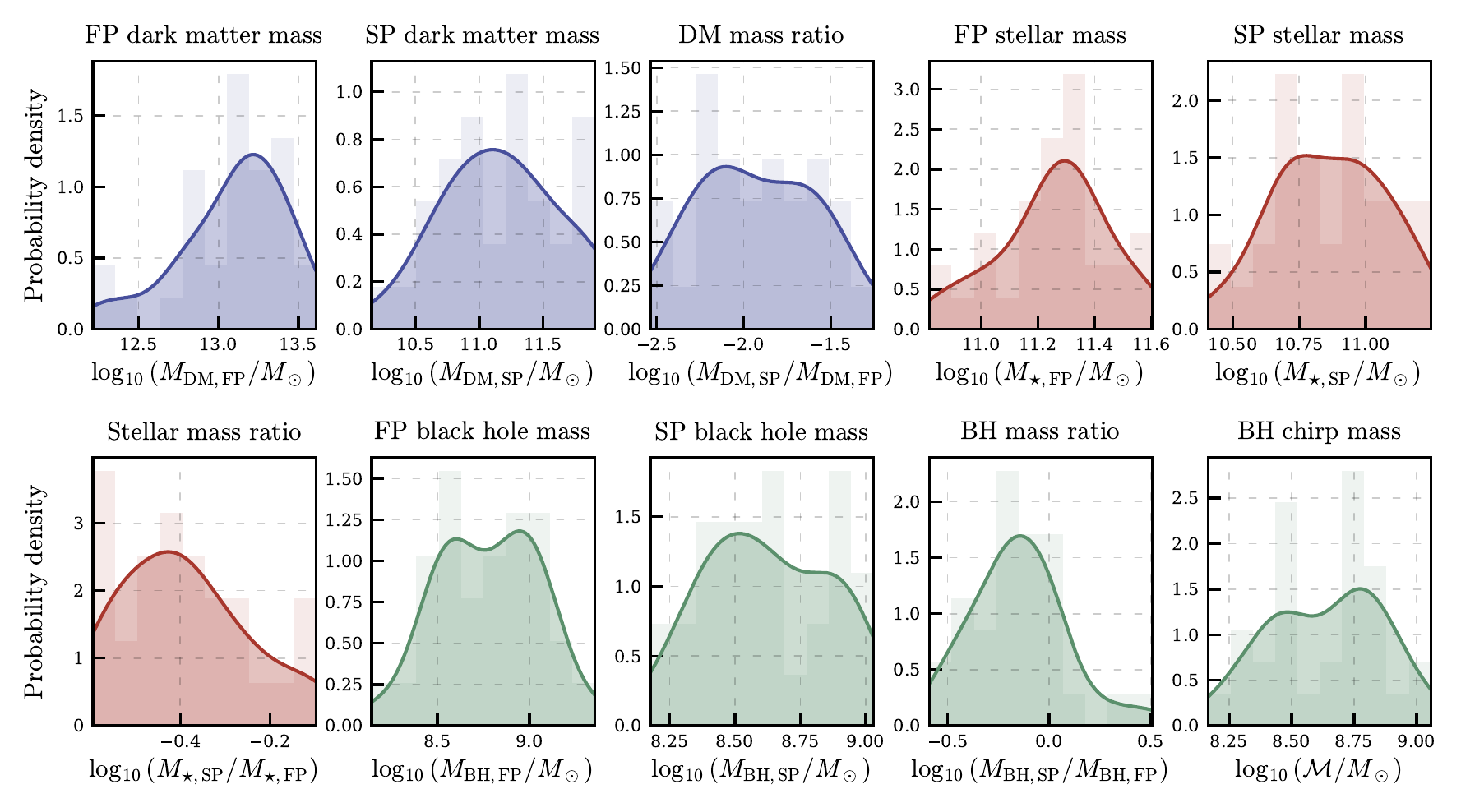}
    \caption{Probability density distributions of the masses of the selected \textsc{IllustrisTNG} galaxy mergers. Blue histograms show the dark matter (DM) masses of the first progenitor (FP), the second progenitor (SP), and their dark matter mass ratio. Red histograms show the corresponding stellar ($\star$) masses and stellar mass ratio, while green histograms show the black hole (BH) masses of the central black holes in the FP and SP, together with their mass ratio and the binary chirp mass. Although the merger sample is selected to have a stellar mass ratio larger than 1/4, the corresponding dark matter mass ratios is much smaller ($\sim 0.003$–$0.06$).}
    \label{fig:massHistoTNG}
\end{figure*}

We present the statistics of our selected sample of \textsc{IllustrisTNG} galaxy mergers. Figure~\ref{fig:massHistoTNG} shows the probability density distributions, computed using kernel density estimation, of the masses of the selected systems, including the stellar, dark matter, and black hole masses. We also report the corresponding mass ratios, as well as the black hole chirp mass for each merger.

The resulting chirp mass distribution lies in the mass range expected to contribute most efficiently to the GWB, with the dominant contribution anticipated from binaries with ${\cal M} > 10^{8}\ M_\odot$ \citep[][]{IzquierdoVillalba2022}.

An interesting feature of our sample is that, although we explicitly select major mergers based on the stellar mass ratio ($q_\star > 1/4$), the corresponding dark matter mass ratios, $q_{\rm DM}=M_{\rm DM,SP}/M_{\rm DM,FP}$ (where $M_{\rm DM,FP}$ and $M_{\rm DM,SP}$ denote the dark matter masses of the first and second progenitors, respectively), are significantly smaller, spanning the range $\sim 0.003$–$0.06$. This reflects the fact that the dark matter haloes of the secondary progenitors have undergone substantial stripping due to their cosmological environment prior to the merger. This behaviour contrasts with the initial conditions commonly adopted in idealised merger simulations, where galaxies are often constructed with identical dark matter haloes or obtained by simply rescaling the secondary progenitor relative to the primary.

\begin{figure*}
	\centering
	\includegraphics[width=0.9\textwidth]{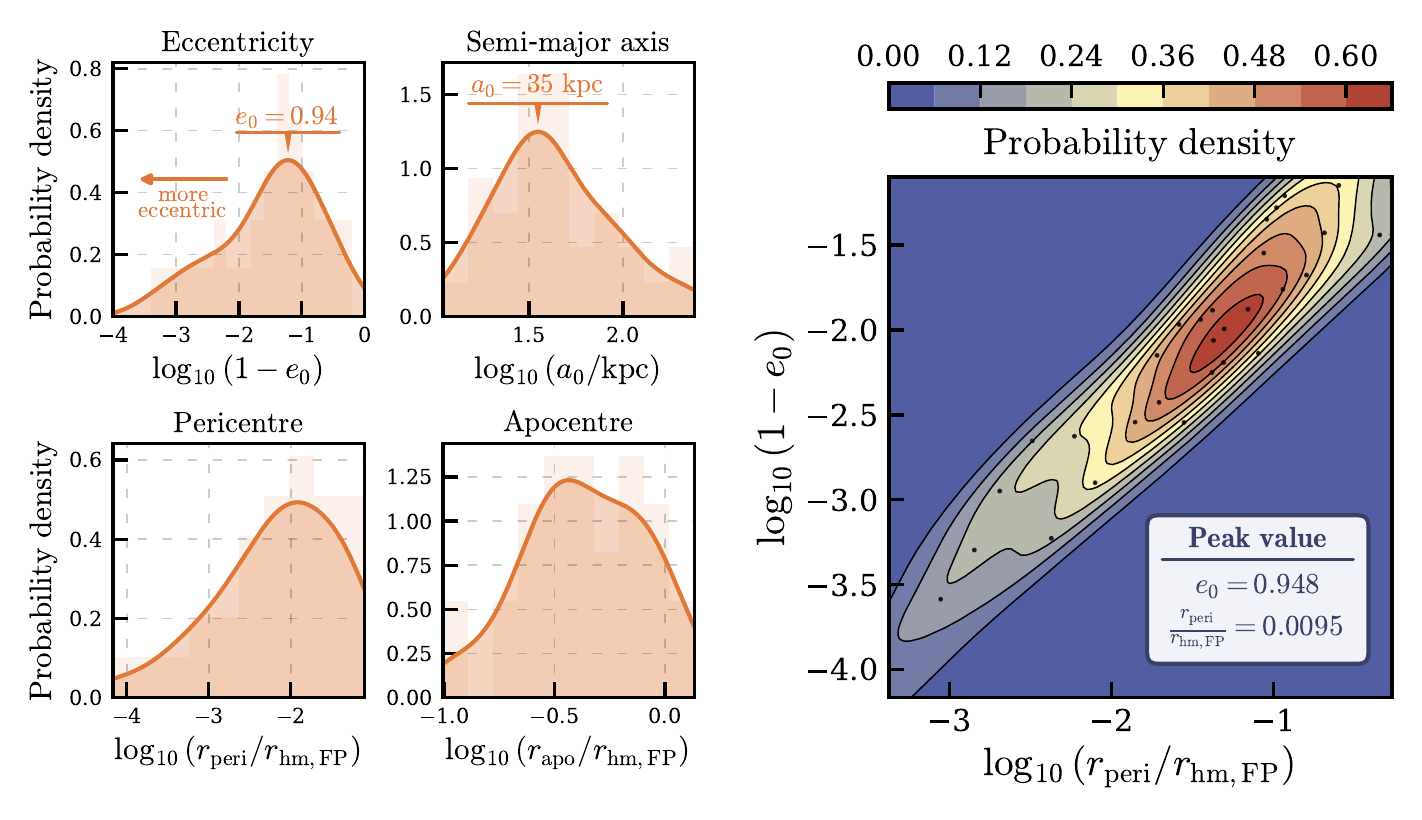}
    \caption{Probability density distributions of the orbital parameters of the selected \textsc{IllustrisTNG} encounters. The left panel shows one-dimensional probability density histograms of the orbital eccentricity, semi-major axis, pericentre, and apocentre, with the latter two normalised by the half-mass radius of the first progenitor ($r_{\rm hm,FP}$). The peak values of the eccentricity and semi-major axis distributions are indicated. The right panel shows the two-dimensional probability density distribution of eccentricity and normalised pericentre, with the peak value highlighted. Overall, cosmological encounters are found to be highly eccentric and to penetrate deeply into the gravitational potential of the first progenitor.}
    \label{fig:orbitHistoTNG}
\end{figure*}

Figure~\ref{fig:orbitHistoTNG} presents the orbital parameters of the selected galaxy encounters. The left-hand panel shows the one-dimensional probability density distributions of the orbital eccentricity, semi-major axis, pericentric distance, and apocentric distance. Cosmological encounters are typically highly eccentric, in agreement with previous studies \citep[e.g.][]{Khochfar2006, Fastidio2024}. The eccentricity distribution peaks at $e_0=0.939$, with a median value of $e_0=0.958$, while the semi-major axis has a peak at $a_0=35$ kpc. As a consequence, the mergers plunge deep into the host potential, with a one-dimensional peak value of pericentric distance normalised by the half-mass radius $r_\mathrm{peri}/r_\mathrm{hm,FP}=0.012$.

The right-hand panel shows the two-dimensional distribution of orbital eccentricity and $r_\mathrm{peri}/r_\mathrm{hm,FP}$. The distribution peaks at $(e_0,\ r_\mathrm{peri}/r_\mathrm{hm,FP}) = (0.948,\ 0.0095)$.

\subsection{Evolution and timescales of black hole binaries}
\label{sec:timescales}

\begin{figure}
	\centering
	\includegraphics[width=\columnwidth]{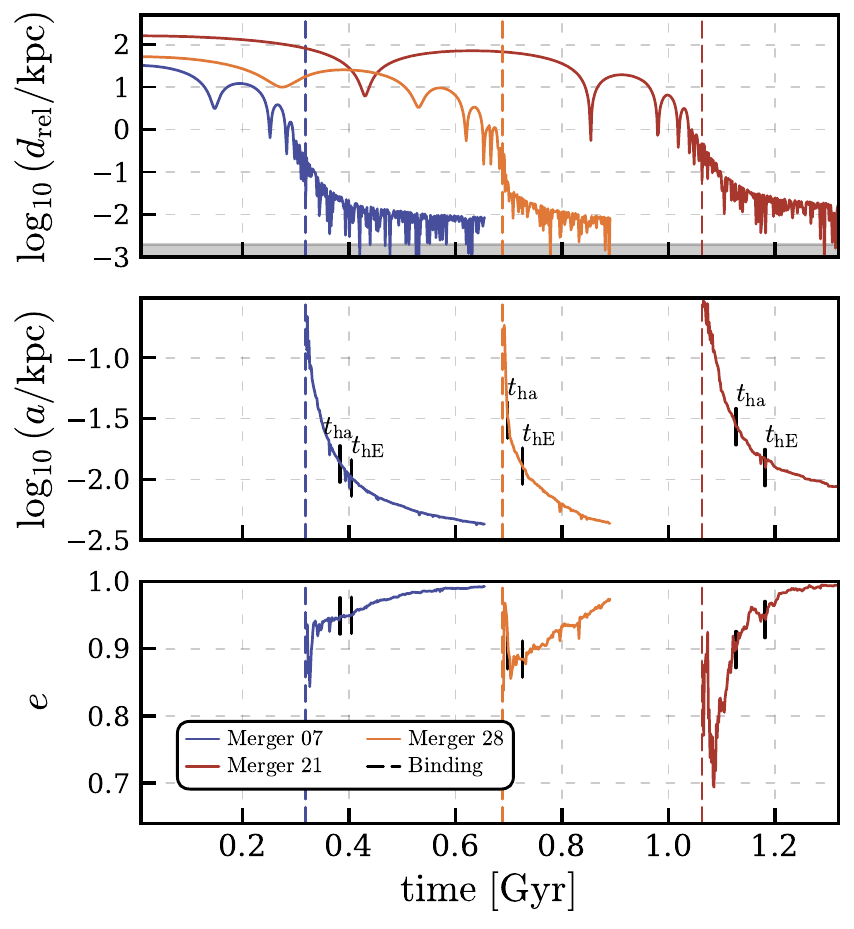}
    \caption{Evolution of three black hole pairs as a function of time, computed with the \textsc{Griffin} $N$-body code. The top panel shows the relative separation, $d_{\rm rel}$, between the two black holes; the shaded region indicates the value of $\varepsilon_{\rm BH,bin}$. The middle panel displays the binary semi-major axis, $a$, while the bottom panel shows the binary eccentricity, $e$. Three representative mergers are shown: merger 07 (blue), merger 21 (red), and merger 28 (orange). The binary binding time ($t_{\rm b}$) is marked by vertical dashed lines for each merger, while the black ticks indicate $t_{\rm ha}$ and $t_{\rm hE}$. The evolution illustrates the progressive inspiral of the black holes from the apocentre of their galactic orbit, followed by energy loss due to dynamical friction during the galaxy merger phase. After binary formation, the subsequent decrease of $a$ and increase of $e$ reflect the extraction of energy and angular momentum driven by stellar hardening.}
    \label{fig:BinEvol}
\end{figure}

Figure~\ref{fig:BinEvol} presents three representative examples of black hole binary evolution. The top panel shows the separation between the two black holes as a function of time, while the middle and bottom panels display the evolution of the semi-major axis and eccentricity, respectively. The three mergers are initialised at the apocentre of the encounter, corresponding to separations of 33, 162, and 52 kpc for mergers 7, 21, and 28, respectively. From these initial conditions, the host galaxies undergo successive pericentric passages and eventually merge, leading to the formation of a bound black hole binary. The binary formation time, $t_{\rm b}$, is indicated by the dashed vertical line and typically occurs after a few pericentric passages. Following its formation, the binary hardens through the ejection of surrounding stars via three-body (slingshot) interactions \citep[][]{Quinlan1996}. This process extracts both energy and angular momentum from the binary, resulting in a gradual decrease of the semi-major axis and a concurrent increase in eccentricity.

\begin{figure}
	\centering
	\includegraphics[width=\columnwidth]{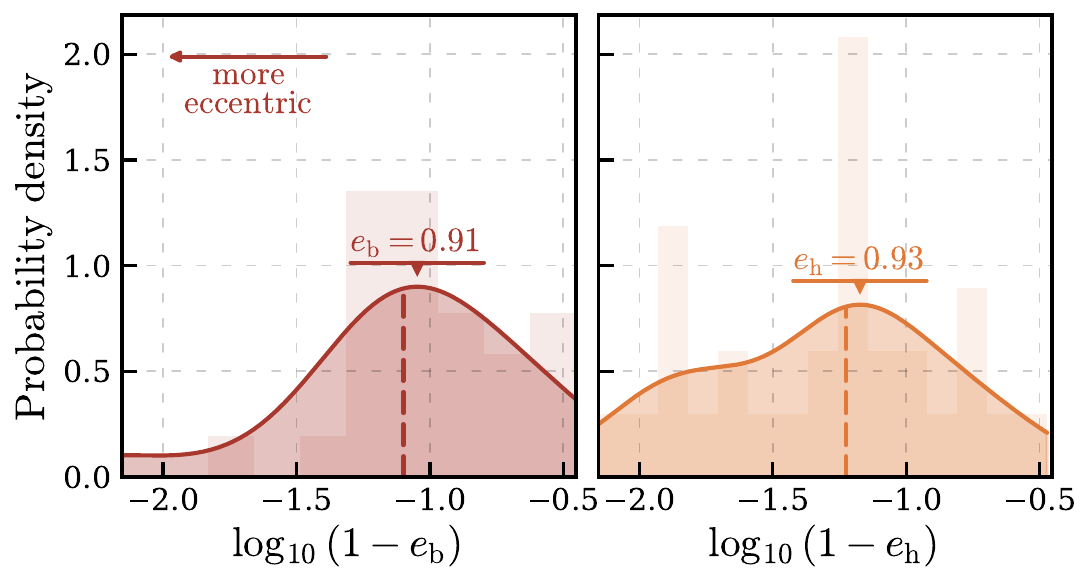}
    \caption{Probability density distributions of binary eccentricities. The left panel shows the distribution of $e_{\rm b}$ at binary formation, while the right panel shows the distribution of $e_{\rm h}$ at the hard separation time. Solid lines indicate kernel density estimates, with their peak values marked, while dashed lines denote the median values. The distributions peak at $e_{\rm b} = 0.91$ and $e_{\rm h} = 0.93$, with corresponding median values of $0.92$ and $0.94$. Binaries are typically formed with high eccentricities and tend to increase their eccentricity during the hardening phase.}
    \label{fig:e_bh}
\end{figure}

Figure~\ref{fig:e_bh} illustrates the probability density distributions of the binary eccentricity at formation, $e_{\rm b}$, and at the hard separation time, $e_{\rm h}$. The binaries are found to be highly eccentric, with peak values of $e_{\rm b}$ and $e_{\rm h}$ of 0.91 and 0.93, respectively, and median values of 0.92 and 0.94, respectively. The distribution of $e_{\rm h}$ exhibits a higher fraction of high-eccentricity systems and fewer low-eccentricity binaries, with a concentration of systems around $\log_{10}(1 - e_{\rm h}) = -2$, corresponding to $e = 0.99$. This shift reflects the systematic increase in eccentricity during the hardening phase. We note that two mergers are not included in this figure, as the galaxies encounter each other on highly eccentric orbits ($e_0 \approx 1$), leading to binaries that exhibit extremely high eccentricities at formation ($\log_{10}(1 - e_{\rm b}) < -2.5$). In these cases, the binaries effectively form already in the hard regime, such that $e_{\rm b}$ is equivalent to $e_{\rm h}$, and they merge almost immediately after formation.

\begin{figure}
	\centering
	\includegraphics[width=\columnwidth]{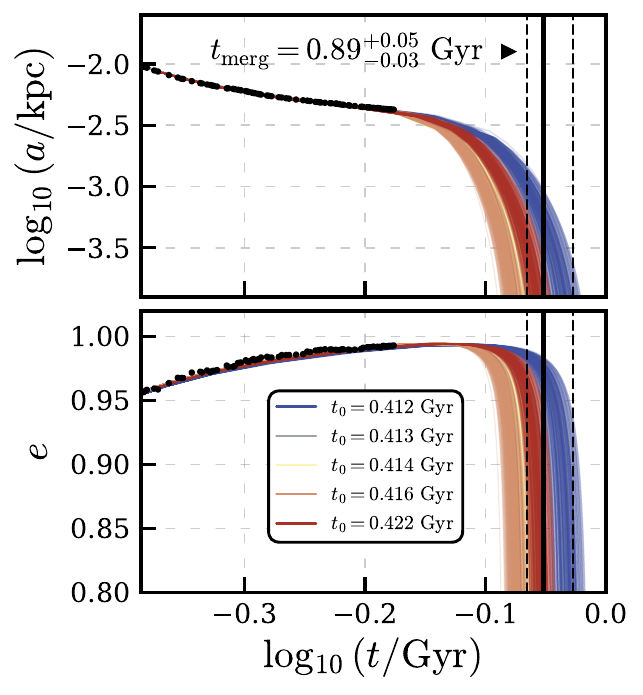}
    \caption{Evolution of the black hole binary in merger 7 as a function of time, computed with the semi-analytic model (SAM) described in Section~\ref{sec:coalescence}. The top panel shows the semi-major axis, $a$, while the bottom panel shows the eccentricity, $e$. Black points correspond to the $N$-body simulation data, and colored lines to SAM integrations. Different colors indicate different initial times for the SAM integration, while multiple curves of the same color represent independent realisations of the $s(t)$ fit drawn from the MCMC posterior distribution. The coalescence time, $t_{\rm coal}$, is defined as the median of the distribution of stopping times and is shown by a solid vertical black line. The 16th and 84th percentiles are indicated by vertical dashed black lines. Overall, the semi-analytic model accurately reproduces the simulation data at early times.}
    \label{fig:BinEvolSAM}
\end{figure}

Figure~\ref{fig:BinEvolSAM} shows the late-time evolution of the binary from merger 7, as modelled with the semi-analytical framework (see equations~\ref{eq:SAM}). The data points correspond to the $N$-body simulation, while the coloured curves represent different realisations of the semi-analytical integrations. The different colours denote distinct choices of initial conditions for the integration, whereas multiple curves of the same colour correspond to different realisations of the MCMC fit of $s(t)$ (see Section~\ref{sec:tmerg} for details). Overall, the semi-analytic model accurately reproduces the simulation data at early times. Here, the integration is terminated when the pericentre falls below $10\, r_{\rm sch}$, which we take to define the coalescence time. The final coalescence time is then estimated as the median of the resulting distribution and is indicated by the black vertical line, with bounds corresponding to the 16th and 84th percentiles.

\begin{figure*}
	\centering
	\includegraphics[width=\textwidth]{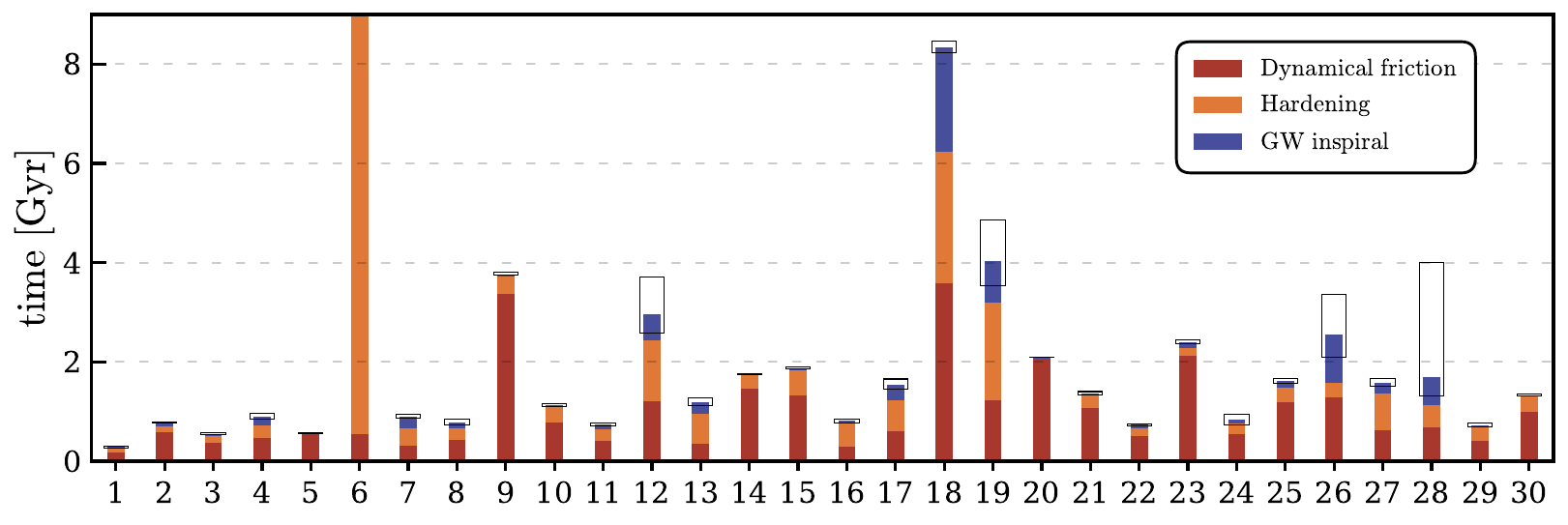}
    \caption{Timescale decomposition of the three phases of binary evolution for the 30 mergers. The red segment represents the duration of the dynamical friction phase, $\Delta t_{\rm df} = t_{\rm b}$, the orange segment the hardening phase, $\Delta t_{\rm h} = t_{\rm gw} - t_{\rm b}$, and the blue segment the gravitational wave driven phase, $\Delta t_{\rm gw} = t_{\rm coal} - t_{\rm gw}$. Black boxes indicate the uncertainty on the coalescence time, defined by the interval between the 16th and 84th percentiles. Merger 6 do not coalesce within the 50 Gyr integration time and instead stall in the hardening phase.}
    \label{fig:timesBar}
\end{figure*}

\begin{figure}
	\centering
	\includegraphics[width=\columnwidth]{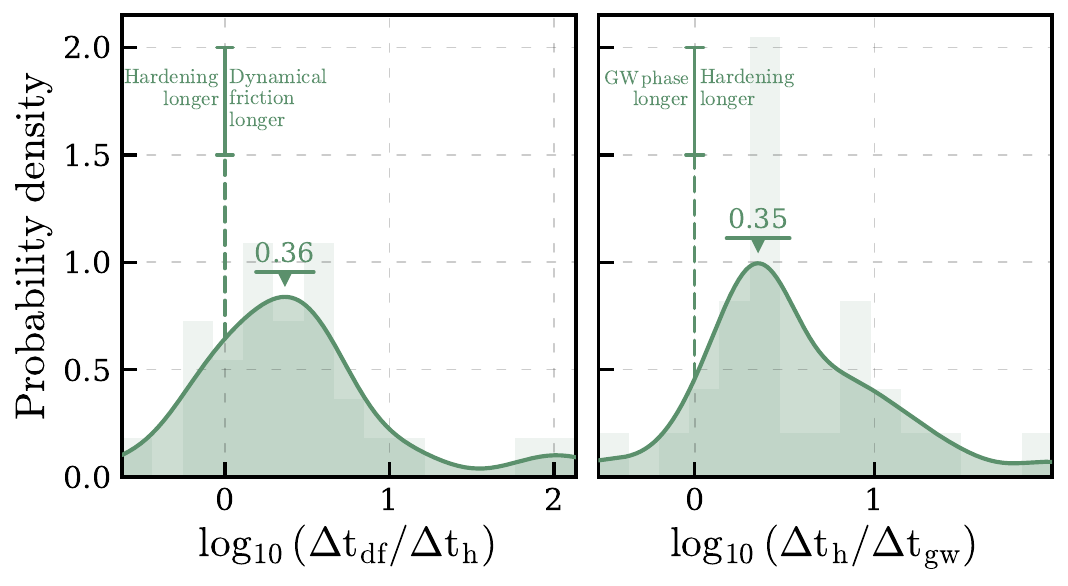}
    \caption{Probability density distributions of timescale ratios. The left panel shows the distribution of $\Delta t_{\rm df} / \Delta t_{\rm h}$, while the right panel shows $\Delta t_{\rm h} / \Delta t_{\rm gw}$. Peak values of the kernel density estimates are indicated. These distributions show that the dynamical friction phase is typically the longest, with a characteristic duration about twice that of the hardening phase, which itself is on average about twice as long as the gravitational wave dominated phase.}
    \label{fig:tscaleRatios}
\end{figure}

Figure~\ref{fig:timesBar} summarises the time spent in each phase of the merger for all 30 systems. Most mergers are completed within $\sim 1$ Gyr (see the distribution of coalescence times and peak value in Fig.~\ref{fig:densityTmerg}), with the bulk of the evolution occurring during the dynamical friction phase. Merger 6, however, does not coalesce within the 50 Gyr integration time and instead stalls in the hardening phase. To quantify the relative importance of each stage, Fig.~\ref{fig:tscaleRatios} presents the probability density distributions of the ratios between the timescales associated with the different phases. The left-hand panel shows the ratio of the time spent in the dynamical friction phase to that spent in the hardening phase. We find that the dynamical friction phase typically lasts about twice as long as the hardening phase. The right-hand panel displays the ratio between the hardening phase and the relativistic regime, where GW emission dominates the evolution. Here again, the hardening phase is typically a factor of 2 longer than the final GW phase.

\subsection{Evolution of binaries within the PTA band}
\label{sec:PTAband}

\begin{figure}
	\centering
	\includegraphics[width=\columnwidth]{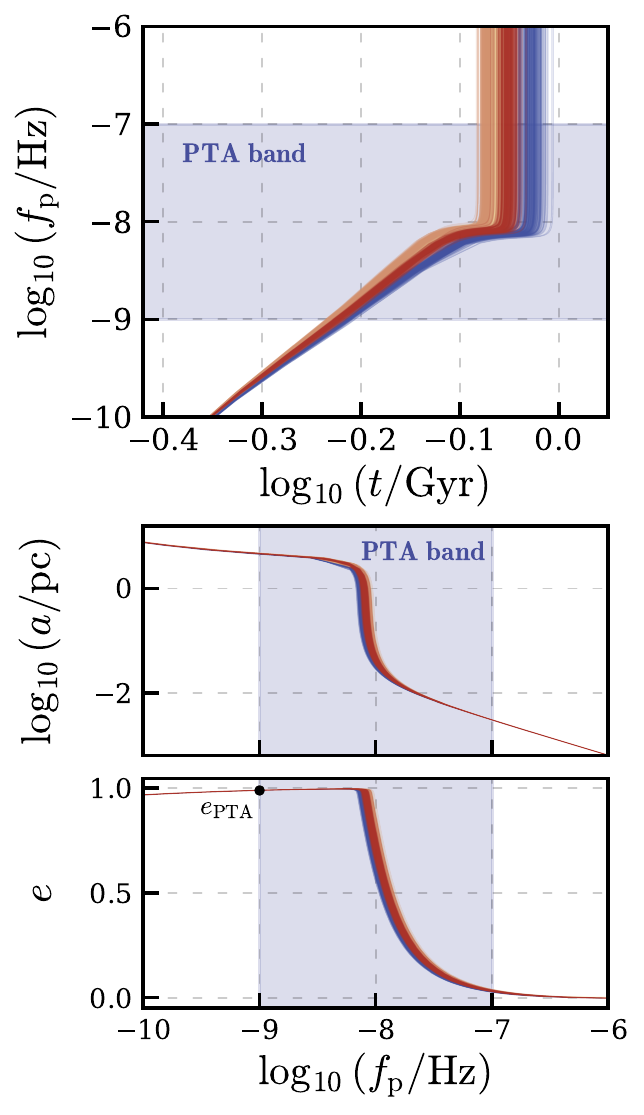}
    \caption{Evolution of the binary from merger 7 within the PTA band. The top panel shows the peak gravitational-wave frequency, $f_{\rm p}$, as a function of time. The middle and bottom panels show the evolution of the semi-major axis, $a$, and eccentricity, $e$, respectively, as functions of $f_{\rm p}$. In all panels, the PTA band ($10^{-9}-10^{-7}$ Hz) is highlighted by the shaded blue region. Different colours correspond to different initial times for the SAM integration (following the same colour scheme as in Fig.~\ref{fig:BinEvolSAM}), while multiple curves of the same colour represent independent realisations of the $s(t)$ fit drawn from the MCMC posterior distribution. The binary spends $302$ Myr within the PTA band and enters it with a very high eccentricity, $e_{\rm PTA} = 0.990$.}
    \label{fig:PTAfreq}
\end{figure}

Figure~\ref{fig:PTAfreq} illustrates the peak gravitational wave frequency, $f_{\rm p}$, emitted by merger 7, defined as \citep[][]{Wen2003}
\begin{equation}
f_{\rm p} = \frac{1}{\pi} \sqrt{\frac{GM}{[a(1 - e^2)]^3}} (1 + e)^{1.1954}.
\end{equation}
The top panel shows the evolution of the peak frequency as a function of time, highlighting the crossing of the pulsar timing array (PTA) band, defined here as the frequency range $10^{-9}-10^{-7}$ Hz. The binary spends approximately $302$ Myr within this frequency range. The bottom panel presents the evolution of the semi-major axis and eccentricity as functions of the peak frequency. We also indicate $e_{\rm PTA}$, the eccentricity at which the binary enters the PTA band, which is found to be $e_{\rm PTA} = 0.990$ in this case.

\begin{figure}
	\centering
	\includegraphics[width=\columnwidth]{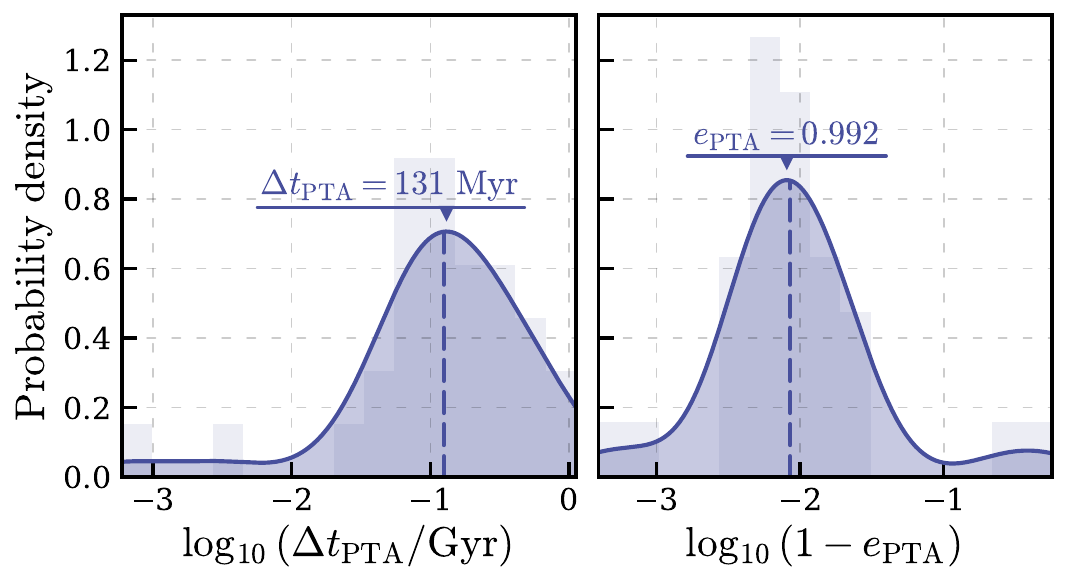}
    \caption{Probability density distributions of the time spent in the PTA band, $\Delta t_{\rm PTA}$ (left), and the eccentricity at entry into the PTA band, $e_{\rm PTA}$ (right). Solid lines show kernel density estimates, with peak values indicated. These distributions show that binaries typically enter the PTA band with very high eccentricities, peaking at $e_{\rm PTA} = 0.992$, and remain in the band for $\sim 131$ Myr. The median residence time and entry eccentricity, indicated by dashed lines, are $125\,\mathrm{Myr}$ and $0.990$, respectively.}
    \label{fig:histoPTA}
\end{figure}

Figure~\ref{fig:histoPTA} presents the probability density functions of the time spent in the PTA band and of the eccentricity at the time of entry into the band for our 30 simulations. The distribution of residence times peaks at $131$ Myr, while the eccentricity at entry peaks at $e_{\rm PTA} = 0.992$. These results indicate that binaries typically enter the PTA band with very high eccentricities and their peak frequencies lie within the $10^{-9}-10^{-7}$ Hz range for several hundred Myr.

\subsection{Scaling relations}

In this section, we leverage our suite of $N$-body simulations to derive empirical relations linking the evolutionary timescales of SMBHBs, namely the dynamical friction phase, the hardening phase, and the total coalescence time, to host galaxy and encounter properties. We begin by introducing the host galaxy properties and examining correlations between these quantities, before investigating how each timescale depends on them.

\subsubsection{Description of the parametres}
\label{sec:params}

The primary objective of this analysis is to relate the BH timescales to quantities that can be robustly measured in cosmological simulations, with the aim of implementing these relations in subgrid models for black hole binary evolution. However, we also explore alternative scaling relations involving less readily accessible parameters. While such quantities may not be directly available in cosmological contexts, they provide additional insight into the underlying physical processes by probing a broader set of galaxy properties than those typically measurable.

Many of the parameters considered in our analysis exhibit significant degeneracies. We describe in Section~\ref{sec:methodFit} how these are handled in order to ensure that the selected relations involve predictors that are as weakly correlated as possible. In some cases, different parameters trace similar physical quantities (e.g. densities measured at different characteristic radii). This redundancy is intentional, as it allows for flexibility in selecting the most informative representation within a family of physically related variables.

Initially, we restricted our analysis to a subset of theoretically motivated features and those showing the strongest individual correlations with the target variables. However, we found that this approach can exclude parameters that, while not dominant at first order, provide important complementary information when combined with primary predictors. Here are the parameters considered:

\begin{itemize}
  \item[$\blacksquare$]  \textbf{Scale lengths.} $\quad$ The half-mass radius of the first progenitor, $r_{\rm hm}$, and the binary influence radius, $r_{\rm inf}$. All distances are expressed in $\mathrm{kpc}$.\\
  
  \item[$\blacksquare$] \textbf{Masses.} $\quad$ We consider the dark matter masses of the first and second progenitors, $M_{\rm DM,FP}$ and $M_{\rm DM,SP}$, their stellar masses, $M_{\rm \star,FP}$ and $M_{\rm \star,SP}$, and their total masses (dark matter plus stars), $M_{\rm tot,FP}$ and $M_{\rm tot,SP}$. We also include the black hole masses, $M_{\rm BH,FP}$ and $M_{\rm BH,SP}$, together with the corresponding mass ratios, $q_{\rm DM}$, $q_{\star}$, $q_{\rm tot}$ and $q_{\rm BH}$.
  We further define the binary mass as $M_{\rm bin}=M_{\rm BH,FP}+M_{\rm BH,SP}$. In addition, we compute initial stellar enclosed masses within the binary influence radius and within $5\,r_{\rm inf}$ for both progenitors, $M_{\rm enc,FP}(r_{\rm inf})$, $M_{\rm enc,SP}(r_{\rm inf})$, $M_{\rm enc, FP}(5\, r_{\rm inf})$ and $M_{\rm enc, SP}(5\, r_{\rm inf})$. We also measure the total (dark matter plus stars) mass enclosed within the initial separation, $M_{\rm enc, tot}(d_{\rm ini})$.
  Finally, for the remnant system (i.e. particles from both progenitors at $t = t_{\rm b}$), we compute the stellar mass enclosed within $r_{\rm inf}$ and $5\,r_{\rm inf}$, denoted $M_{\rm enc, rem}(r_{\rm inf})$ and $M_{\rm enc, rem}(5\,r_{\rm inf})$. All masses are expressed in $M_\odot$.\\
  
  \item[$\blacksquare$] \textbf{Densities.} $\quad$  We compute the stellar density of both progenitors at $r_{\rm inf}$, $5\,r_{\rm inf}$, and $r_{\rm hm}$, i.e. $\rho_{\star,\rm FP}$ and $\rho_{\star,\rm SP}$ evaluated at these radii. These quantities are computed both at the beginning of the simulation and at the binary formation time.
  For the remnant galaxy (particles from both progenitors at $t = t_{\rm b}$), we similarly compute the stellar density at $r_{\rm inf}$, $5\,r_{\rm inf}$, and $r_{\rm hm}$, denoted $\rho_{\star,\rm rem}(r_{\rm inf})$, $\rho_{\star,\rm rem}(5\,r_{\rm inf})$, and $\rho_{\star,\rm rem}(r_{\rm hm})$. We also look at the merger-induced variation in the central stellar density of the first progenitor, $\Delta \rho_{\rm FP}(r_{\rm inf})$. All densities are expressed in $M_\odot\,\mathrm{kpc}^{-3}$.\\
  
  \item[$\blacksquare$]  \textbf{Orbit.} $\quad$ We include the initial orbital parameters of the galaxy merger, namely the semi-major axis $a_0$, eccentricity $e_0$, pericentre $r_{\rm peri}$, and apocentre $r_{\rm apo}$. We also define an interpenetration parameter, $r_{\rm inter} = r_{\rm hm}/r_{\rm peri}$.
  For the black hole binary, we consider the semi-major axis, eccentricity, and pericentre at formation, $a_{\rm b}$, $e_{\rm b}$, and $r_{\rm peri,b}$, as well as at the hard separation time, $a_{\rm h}$, $e_{\rm h}$, and $r_{\rm peri,h}$.\\

  \item[$\blacksquare$] \textbf{Hardening parametres.} $\quad$ We include the median hardening rate\footnote{Median value of $s(t)$ computed over time bins, see Section~\ref{sec:coalescence}.}, $\tilde{s}$, and the eccentricity growth parameter, $K$. \\

  \item[$\blacksquare$] \textbf{Kinematics.} $\quad$ We compute the 1D stellar velocity dispersion of the remnant (at $t = t_{\rm b}$) measured within $r_{\rm inf}$, $5\, r_{\rm inf}$ and $r_{\rm hm}$, $\sigma_{\rm inf}$, $\sigma_{\rm 5inf}$ and $\sigma_{\rm hm}$, respectively, as well as a measure of rotation, $\bar{v}_{\varphi}/\sigma_{\varphi}$, where $\bar{v}_{\varphi}$ is the mean stellar rotation velocity of the remnant and $\sigma_{\varphi}$ the velocity dispersion in the azimuthal direction, also computed at $r_{\rm inf}$, $5\, r_{\rm inf}$ and $r_{\rm hm}$. We finaly compute the anisotropy parametre at the same radii $\beta_{\rm inf}$, $\beta_{\rm 5inf}$, $\beta_{\rm hm}$. All velocities are expressed in $\mathrm{km}/\mathrm{s}$.
  
\end{itemize}

\subsubsection{Method for finding the relations}
\label{sec:methodFit}

Many methods can be used to relate host galaxy properties to black hole binary coalescence timescales, ranging from regularised linear models such as the lasso \citep[][]{Tibshirani1996}, to tree based ensembles \citep[][]{Breiman2001, Friedman2001}, Gaussian process emulators \citep[][]{Kimeldorf1970, Rasmussen2006}, nonlinear dimensionality reduction methods such as kernel PCA \citep[][]{Scholkopf1998}, and topological data analysis \citep[][]{Chazal2021}. In our case, however, the training set is small, only 30 high resolution simulations, which makes highly flexible approaches difficult to constrain robustly and, in practice, often less transparent to interpret physically. Following the same general logic as \citet{HolleyBockelmann2025}, we therefore favor a simple, empirical strategy based on linear regression and best-subset selection, because it remains usable with limited data, preserves interpretability, and yields a compact scaling relation that can be readily implemented in subgrid models for binary BH coalescence times.

We first apply a logarithmic transformation to the parametres in order to capture power law dependencies. The data are then standardised by subtracting the mean and dividing by the standard deviation of each feature. To identify candidate scaling relations involving $n$ features, we perform an exhaustive search over all possible combinations of $n$ predictors drawn from the full set of $n_{\rm tot}$ parameters ${n_{\rm tot} \choose n}$. For each combination, we fit a linear model to the data and evaluate its goodness of fit using the coefficient of determination\footnote{The coefficient of determination quantifies the fraction of the variance in the dependent variable that is explained by the regression model, with values ranging from 0 (no explanatory power) to 1 (perfect fit).}, $R^2$, defined as
\begin{equation}
    R^2 = 1-\frac{\sum_{i=1}^{k}(y_i-f_i)^2}{\sum_{i=1}^{k}(y_i-\overline{y})^2},
\end{equation}
where $k$ is the number of data points, $y_i$ the true values, $f_i$ the model predictions, and $\overline{y}$ the sample mean.

To avoid selecting relations that involve strongly correlated predictors, we quantify multicollinearity using the variance inflation factor (VIF). For each variable $X_i$ included in a given model, the VIF is defined as
\begin{equation}
    \mathrm{VIF}_i = \frac{1}{1-R_i^2},
\end{equation}
where $R_i^2$ is the coefficient of determination obtained by regressing $X_i$ against all other predictors in the model. The VIF therefore measures how much the variance of the estimated regression coefficient is inflated due to linear dependencies with the other variables. For a relation involving $n$ predictors, we compute the VIF for each variable and retain the maximum value as a proxy for the overall level of multicollinearity in the model. This criterion is then used to penalise and discard candidate relations with strongly correlated features. In practice, we discard all candidate relations with a maximum VIF greater than 5.

To further assess the robustness and predictive power of the selected relations, we perform a leave-one-out cross validation (LOOCV). For each candidate model, we iteratively exclude one data point from the sample, fit the model on the remaining $k-1$ points, and evaluate the prediction error on the excluded point. Repeating this procedure for all data points yields a distribution of prediction errors, from which we compute the root mean square error (RMSE). The final scaling relation is then chosen as the model that simultaneously achieves a high $R^2$, a low maximum VIF, and minimises the LOOCV RMSE.

\subsubsection{Scaling relations for dynamical friction phase}
\label{sec:DFphase}

\begin{figure*}
	\centering
	\includegraphics[width=\textwidth]{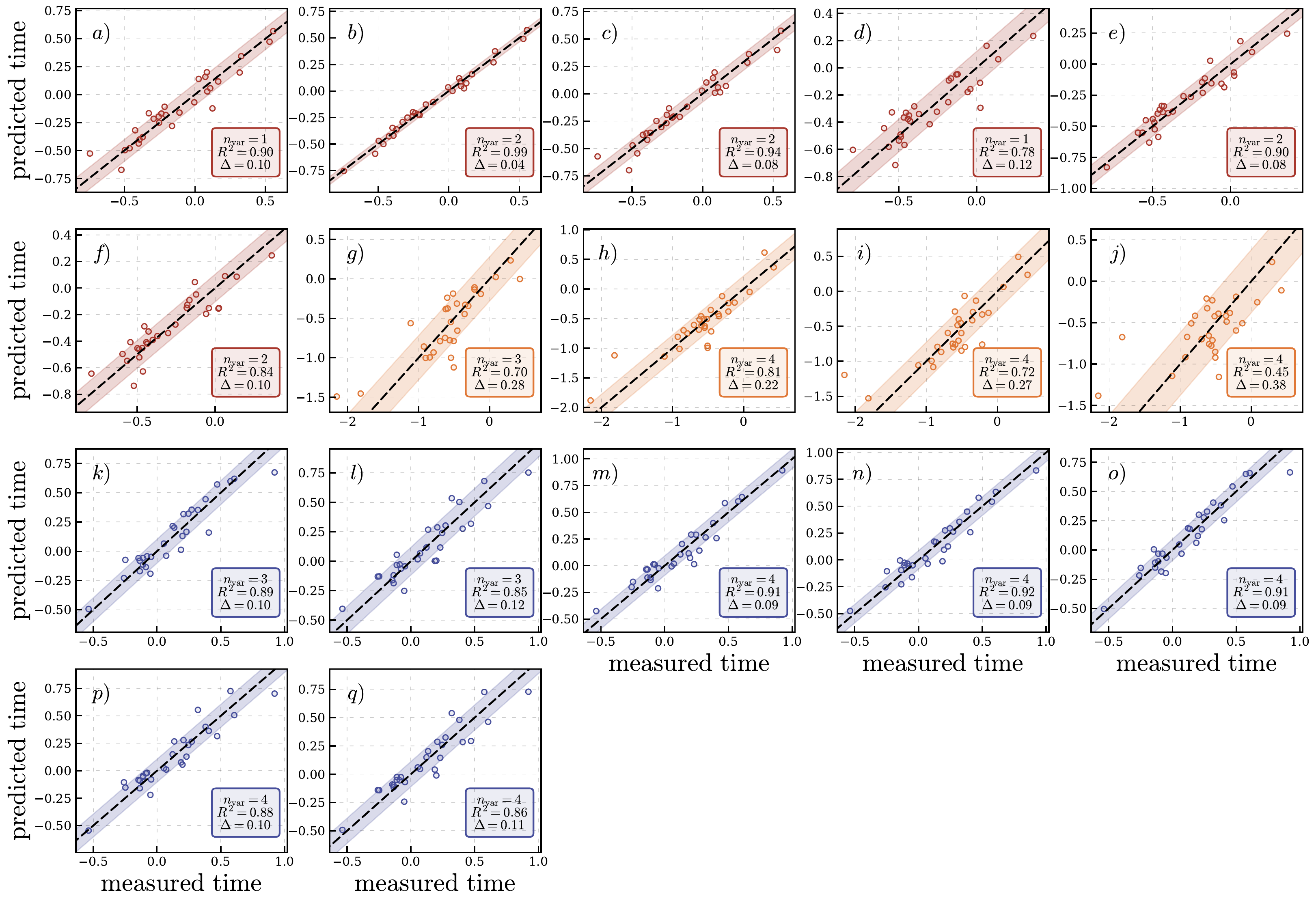}
    \caption{Measured versus predicted timescales for the scaling relations listed in Table~\ref{tab:relations}. Each panel corresponds to the relation of matching letter in Table~\ref{tab:relations}. Red, orange, and blue panels correspond to the dynamical friction ($\Delta t_{\rm df}$), hardening ($\Delta t_{\rm h}$), and coalescence ($t_{\rm coal}$) timescales, respectively. All quantities are plotted as $\log_{10}$ values in units of Gyr. The dashed line shows the one-to-one relation, and the shaded band indicates the intrinsic scatter $\Delta$ of each relation. The number of predictors $n_{\rm var}$, the coefficient of determination $R^2$, and the intrinsic scatter $\Delta$ are reported in each panel.}
    \label{fig:allRelations}
\end{figure*}

We apply the procedure described above to identify scaling relations between the time spent in the dynamical friction phase (from the start of the simulation to $t_{\rm b}$) and the host galaxy and encounter parameters introduced in Section~\ref{sec:params}. We first search for a single-parameter relation in order to identify the most informative predictor. This yields relation (\ref{rel:a}) in Table~\ref{tab:relations}. Panel (a) of Fig.~\ref{fig:allRelations} shows the predicted dynamical friction timescale from this one-parameter relation compared to the measured values. We find that the dynamical friction time is already well captured by a single parameter, with a coefficient of determination $R^2 = \mathrm{0.90}$ and a scatter of $\Delta = \mathrm{0.10}$. The most informative parameter is the semi-major axis of the galaxy encounter ($a_0$).

We then extend the analysis to two-parameter relations in order to improve the predictive power. This leads to relation (\ref{rel:b}) in Table~\ref{tab:relations}, with the corresponding predicted versus measured values shown in panel (b) of Fig.~\ref{fig:allRelations}. The parameter that most effectively complements the semi-major axis is the velocity dispersion within the half-mass radius ($\sigma_{\rm hm}$). This two-parameter relation yields a very high coefficient of determination, $R^2 = \mathrm{0.99}$, together with a low scatter of $\Delta=\mathrm{0.04}$. However, this quantity may not be straightforward to compute in practice, as it requires access to the properties of the remnant galaxy after the merger. We therefore provide an alternative two-parameter relation with nearly identical predictive power, relation (\ref{rel:c}), which instead depends on the black hole mass of the first progenitor ($M_{\rm BH, FP}$). This relation also achieves a very high coefficient of determination, $R^2 = \mathrm{0.94}$, together with a low scatter of $\Delta=\mathrm{0.08}$.

Since our simulations are initialised at apocentre, this setup tends to enhance correlations with the orbital properties of the encounter, as systems with larger apocentric distances require more time to reach their first pericentric passage. To assess the impact of this choice, we repeat the analysis by redefining the time origin at a fixed separation, $d_{0,\rm std} = 30$ kpc, common to all mergers. The resulting one- and two-parameter relations are reported as relations (\ref{rel:d}), (\ref{rel:e}) and (\ref{rel:f}) in Table~\ref{tab:relations}, with the corresponding predicted versus measured comparisons shown in panels (d), (e) and (f) of Fig.~\ref{fig:allRelations}. We find that, despite a slight increase in scatter and decrease in $R^2$, the form of the relations remains largely unchanged.

These results indicate that the orbital configuration of the galaxy encounter plays a dominant role in setting the duration of the dynamical friction phase. Given that this phase typically represents the longest stage of the evolution (see Fig.~\ref{fig:tscaleRatios}), the total black hole coalescence time is expected to be strongly influenced by the cosmological orbital parameters.

\subsubsection{Scaling relations for hardening phase}
\label{sec:HardeningPhase}

\begin{figure}
	\centering
	\includegraphics[width=\columnwidth]{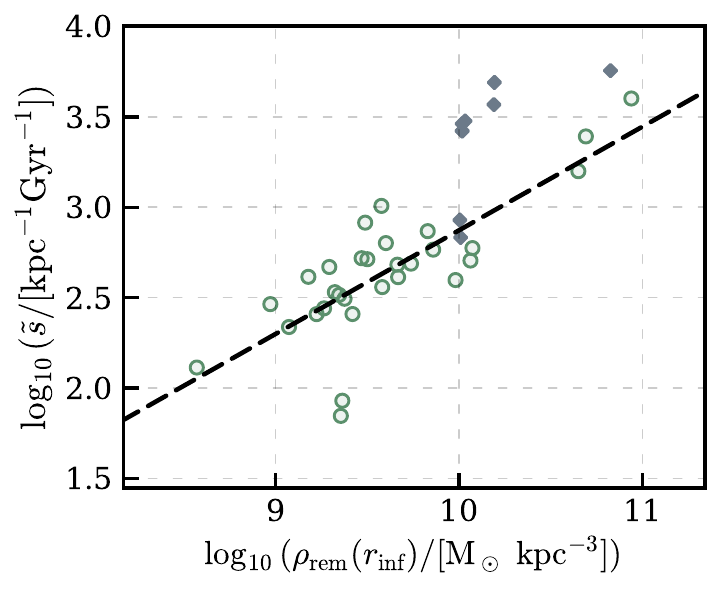}
    \caption{Correlation between the hardening rate $\tilde{s}$ and the stellar density of the merger remnant evaluated at the binary's influence radius, $\rho_{\rm rem}(r_{\rm inf})$. Open circles represent our simulations, and grey diamonds show data from \citet{HolleyBockelmann2025} that overlap with the parameter space explored here. The dashed line indicates the best-fit relation to our data.}
    \label{fig:rhoS}
\end{figure}

Figure~\ref{fig:rhoS} shows the correlation between the hardening rate ($\tilde{s}$) and the central stellar density of the remnant ($\rho_{\rm rem}(r_{\rm inf})$)\footnote{We also find a very similar correlation between $\tilde{s}$ and the ratio of the central density to the velocity dispersion, as theoretically expected.}. This provides a useful consistency check, ensuring that the hardening process is physically well captured in our simulations. The grey diamond markers correspond to the data from \citet{HolleyBockelmann2025}, which overlap with the region of the $(s, \rho)$ parameter space spanned by our simulations, for comparison.

We first search for scaling relations for the hardening phase (from $t_{\rm b}$ to $t_{\rm gw}$) by allowing the full set of parameters, including those not directly accessible in cosmological simulations, in order to identify the most informative predictors. The best-fitting relation ($R^2=0.80$) formally involves the hardening rate, as well as the orbital properties of the hard binary through the pericentre $r_{\rm peri,h}$ and the semi-major axis $a_{\rm h}$. This relation is not directly usable for predictive purposes, as the hardening rate is better interpreted as an outcome of the binary evolution rather than an input parameter. For this reason, we do not report it in Table~\ref{tab:relations}. Nevertheless, it provides useful insight into the underlying physical mechanisms and allows for a direct connection with theoretical expectations for the hardening process. In this relation, the coefficients associated with the hardening rate and the semi-major axis are negative, while that associated with the pericentre is positive. This behaviour can be understood by considering first the idealised case in which the evolution is driven solely by stellar hardening. In this limit, the characteristic hardening timescale is given by
\begin{equation}
    \left.\Delta t_{\rm h}\right|_{\star} \sim \left| \frac{a}{\dot a} \right| = \frac{1}{sa}
\end{equation}
where $\left.\Delta t_{\rm h}\right|_{\star}$ denotes the contribution to the hardening timescale arising purely from the hardening phenomenon and the last equality follows from equation~(\ref{eq:dade}). The negative dependence on both the hardening rate and the semi-major axis directly arises from this equation. The inclusion of the pericentre further captures the role of orbital geometry in the subsequent evolution: small pericentre passages can efficiently enhance energy losses through gravitational wave emission, thereby accelerating the transition from the hardening phase to the gravitational wave–dominated regime, which accounts for the positive coefficient associated with $r_{\rm peri,h}$.

If this relation involving the hardening rate is excluded, we obtain the three-parameter relation (\ref{rel:g})\footnote{We do find a relation with the central density of the remnant as an equivalent of the hardening rate, consistent with the correlation in Fig.~\ref{fig:rhoS}, nevertheless, the corresponding $R^2$ is slightly lower than that of relation (\ref{rel:g}).} in Table~\ref{tab:relations}. The corresponding comparison between predicted and measured values is shown in panel (g) of Fig.~\ref{fig:allRelations}. This relation yields a coefficient of determination of $R^2=0.70$ and a scatter of $\Delta=0.28$. It retains the dependence on $r_{\rm peri,h}$ and $a_{\rm h}$, while replacing the hardening rate with the binary black hole mass $M_{\rm bin}$. The positive coefficient associated with $M_{\rm bin}$ indicates that more massive binaries experience longer hardening timescales. The observed dependence of the hardening timescale on the binary mass indicates a departure from the ideal steady-state regime of \citep[][]{Quinlan1996}, in which the hardening timescale is expected to be independent of mass.

Relation (\ref{rel:h}) corresponds to a four-parameter scaling for the hardening phase. As in relation (\ref{rel:g}), it retains a dependence on $M_{\rm bin}$ and $a_{\rm h}$, but replaces $r_{\rm peri,h}$ with the binary eccentricity at formation, $e_{\rm b}$ (we note that alternative relations retaining $r_{\rm peri,h}$ can also be found, albeit with slightly lower $R^2$ values), and includes the velocity dispersion measured at the centre of the remnant, $\sigma_{\rm 5inf}$. The inclusion of this additional parameter improves the coefficient of determination to $R^2=0.81$, with an associated scatter of $\Delta=0.22$.

From the set of relations identified so far for the hardening phase, a consistent picture emerges. When no restriction is imposed on the parameter space, we find that the key quantities governing the evolution are two parameters characterizing the binary orbit, the binary mass, and the central properties of the remnant, namely its central density and velocity dispersion. These dependencies are broadly consistent with theoretical expectations \citep[][]{Quinlan1996}. However, a practical limitation is that such quantities are not readily accessible, as they typically require high-resolution simulations to be robustly determined.

We now progressively restrict the accessible parameter space in order to identify more practical relations while retaining some predictive power. Relation (\ref{rel:i}) corresponds to a four-parameter scaling obtained when excluding orbital parameters at $t_{\rm h}$. It instead depends on the orbital properties at binary formation, namely the eccentricity $e_{\rm b}$ and semi-major axis $a_{\rm b}$, together with the anisotropy parameter measured within $5\, r_{\rm inf}$ and the influence radius of the binary, $r_{\rm inf}$. This relation yields a coefficient of determination of $R^2=0.72$.

Relation (\ref{rel:j}) represents the best four-parameter scaling obtained when excluding all orbital information about the black hole binary. It exhibits a significantly lower coefficient of determination, $R^2=0.45$, and a relatively large scatter, $\Delta=0.38$. This suggests that the relation should be interpreted with caution, as it is likely affected by limited statistics and potential overfitting. More generally, it highlights the intrinsic difficulty of accurately describing the hardening phase in the absence of direct information on the binary orbit. The hardening phase is sensitive to detailed aspects of the binary dynamics that are not captured by global galaxy properties. Moreover, the stochastic nature of the binary eccentricity at formation \citep[see][]{Rawlings2023, Gualandris2026} introduces an additional source of scatter, which cannot be fully mitigated even with a detailed dynamical description. Nevertheless, the fact that relation (\ref{rel:j}), which relies exclusively on global properties still retains non negligible predictive power is encouraging in the context of estimating the total coalescence time. Combined with the fact that the dynamical friction phase is both well described by these parameters and typically dominates the overall coalescence timescale, this suggests that reasonably accurate scaling relations for the total coalescence time can be constructed despite the limitations discussed above.

\subsubsection{Scaling relations for black holes coalescence time}
\label{sec:tmergRelations}

Relation (\ref{rel:k}) corresponds to the three-parameter scaling relation for the total coalescence time without any restriction on parameter accessibility. It recovers a dependence on the initial semi-major axis of the encounter ($a_0$), which was identified as an excellent predictor of the dynamical friction phase, as well as on $e_{\rm h}$, which provides strong predictive power for the hardening phase. The central velocity dispersion of the remanent ($\sigma_{5\rm inf}$) also enters as a secondary parameter, contributing to both the dynamical friction phase (at larger scales) and the hardening phase (cf. relations \ref{rel:b}, \ref{rel:e} and \ref{rel:h}). Although this relation achieves a high coefficient of determination ($R^2 = 0.89$), it relies on $e_{\rm h}$, which is not readily accessible in lower-resolution cosmological simulations. We therefore provide an alternative relation, (\ref{rel:l}), which is equivalent in form but involves parameters that are more easily measurable or inferable from cosmological simulations. This relation still depends on $a_0$, but replaces $e_{\rm h}$ with the initial orbital eccentricity $e_0$ and includes the central enclosed mass of the second progenitor ($M_{\rm enc,SP}(5\, r_{\rm inf})$). It retains a similarly high predictive power, with $R^2 = 0.85$.

We further explored four-parameter relations in order to improve the predictive performance. Without imposing any restriction on parameter accessibility, we find that $a_0$ and $e_{\rm h}$ consistently emerge as the dominant parameters, appearing in all of the best-fitting relations. These are complemented by the central anisotropy, $\beta_{\rm 5inf}$, together with the central density of the remnant, $\rho_{\rm rem}(5\, r_{\rm inf})$, yielding relation (\ref{rel:m}) with $R^2=0.91$. An even tighter relation is obtained when replacing the density with the central velocity dispersion, $\sigma_{5\rm inf}$, leading to relation (\ref{rel:n}) with $R^2=0.92$.

The apparent interchangeability between $\sigma_{5\rm inf}$ and $\rho_{\rm rem}(5\, r_{\rm inf})$ can be understood from the fact that the velocity dispersion implicitly encodes information about both the central density and the enclosed mass. Indeed, we recover the following relation
\begin{equation}
\begin{split}
    \log_{10}(\sigma_{5\rm inf}/[\mathrm{km}\, \mathrm{s}^{-1}]) ={}& 0.1\, \log_{10}(\rho_{\rm rem}(5\, r_{\rm inf})/[M_\odot\, \mathrm{kpc}^{-3}]) \\
    &+ 0.3\, \log_{10}(M_{\rm enc, rem}(5\, r_{\rm inf})/M_\odot) -4.4
    \label{eq:sigRhoM}
\end{split}
\end{equation}
with a coefficient of determination $R^2=0.90$. This scaling is broadly consistent with expectations from simple virial arguments, in which the velocity dispersion traces both the enclosed mass and the density of the system.\footnote{Under the assumption of a virialized system, $\sigma(r)^2 \sim GM_{\rm enc}(r)/r$, and adopting $\rho(r) \sim M_{\rm enc}(r)/r^3$, one obtains $\sigma(r) \propto M_{\rm enc}(r)^{1/3}\rho(r)^{1/6}$, corresponding to logarithmic slopes of $1/3$ and $1/6$ with respect to $M_{\rm enc}$ and $\rho$, respectively, in reasonable agreement with equation~(\ref{eq:sigRhoM}).}

Alternatively, relation (\ref{rel:o}) links the coalescence time to $a_0$, $e_{\rm h}$, the binary influence radius ($r_{\rm inf}$), and the enclosed mass of the second progenitor ($M_{\rm enc,SP}(5\, r_{\rm inf})$), with a coefficient of determination $R^2 = \mathrm{0.91}$.

Starting from relation (\ref{rel:o}), it is possible to construct corresponding scalings based on quantities accessible in cosmological simulations, provided that the information carried by $e_{\rm h}$ is replaced by that encoded in the initial eccentricity, $e_0$. This leads to relations (\ref{rel:p}) and (\ref{rel:q}) for the total coalescence time. The two relations differ only in the definition of the enclosed mass of the second progenitor: relation (\ref{rel:p}), which provides the better fit ($R^2=0.88$), depends on the mass enclosed within the binary influence radius, $r_{\rm inf}$, while relation (\ref{rel:q}), with a slightly lower coefficient of determination ($R^2=0.86$), instead relies on the mass enclosed within $5\, r_{\rm inf}$. The latter may be preferable in the context of cosmological simulations, where the innermost density structure is often less reliably resolved.

The fact that the coefficients of determination remain comparable between the cases where all parameters are available and those restricted to global, accessible quantities, in both the three- and four-parameter relations, highlights an important difference with the hardening phase: while accurately modelling the hardening timescale requires detailed information on the binary dynamics, the total coalescence time can be predicted with comparable accuracy using only global parameters. This can be understood for two main reasons. First, the dynamical friction phase, which typically dominates the overall coalescence timescale, is very well described by global properties. Secondly, as shown in the previous section, these global parameters still retain partial information on the hardening timescale, allowing them to capture part of the underlying dependence despite the absence of detailed dynamical quantities.

\begin{table*}
    \centering
    \caption{Scaling relations for the binary evolution. Relations \textbf{a}--\textbf{f} model the dynamical friction timescale $\Delta t_{\rm df}$ as a function of host galaxy and encounter parameters. In relations \textbf{a}--\textbf{c}, $\Delta t_{\rm df}$ is measured from the initial apocentre of the galactic encounter to the moment of binary formation $t_{\rm b}$; in relations \textbf{d}--\textbf{f}, it is measured from the time at which the two galaxies are separated by $d_{\rm 0,std}$ to $t_{\rm b}$. Relations \textbf{g}--\textbf{j} model the hardening timescale $\Delta t_{\rm h} = t_{\rm gw} - t_{\rm b}$, and relations \textbf{k}--\textbf{q} model the total coalescence time $t_{\rm coal}$, both as functions of host galaxy and encounter parameters. For each relation we report: the number of predictors $n_{\rm var}$; the coefficient of determination $R^2$; the adjusted coefficient of determination $R^2_{\rm adj}$, which penalises model complexity and allows fair comparison across relations with different numbers of parameters; the intrinsic scatter $\Delta$; the maximum variance inflation factor ($\mathrm{VIF}$), which quantifies multicollinearity among predictors; and the root-mean-square error of a leave-one-out cross-validation ($\mathrm{LOOCV}$), which estimates out-of-sample predictive performance. Relations labelled in bold depend solely on global parameters extractable from cosmological simulations.}
    \label{tab:relations}
    \begin{tabular}{l p{1.6cm} l c c c c c c}
        \hline
         & Relation & & $n_{\rm var}$ & $R^2$ & $R^2_{\rm adj}$ & $\Delta$ & VIF & LOOCV  \\
        \hline
        \multicolumn{9}{c}{\textbf{Dynamical friction}} \\
        \relationbf & $\log_{10}(\Delta t_{\rm df}/\mathrm{Gyr}) $ &$= +0.9273\ \log_{10}(a_0/\mathrm{kpc}) -1.6392$ & 1 & 0.90 & 0.90 & 0.10 & \ding{55} & 0.33 \\
        \relationbf & $\log_{10}(\Delta t_{\rm df}/\mathrm{Gyr}) $ &$= + 1.0112\ \log_{10}(a_0/\mathrm{kpc}) -1.4608\ \log_{10}(\sigma_{\rm hm}/[\mathrm{km\, s^{-1}}]) -2.5777$   & 2 & 0.99 & 0.98 & 0.04 & 1.1 & 0.13 \\
        \relationbf & $\log_{10}(\Delta t_{\rm df}/\mathrm{Gyr}) $ &$= + 1.0746\ \log_{10}(a_0/\mathrm{kpc}) -0.2919\ \log_{10}(M_{\rm BH,FP}/M_\odot)  + 0.6865$   & 2 & 0.94 & 0.94 & 0.08 & 1.6 & 0.27 \\
        \relationbf & $\log_{10}(\Delta t_{\rm df}/\mathrm{Gyr}) $ &$= +0.7104\ \log_{10}(a_0/\mathrm{kpc}) -1.4562$ & 1 & 0.78 & 0.77 & 0.12 & \ding{55} & 0.50 \\
        \relationbf & $\log_{10}(\Delta t_{\rm df}/\mathrm{Gyr}) $ &$= + 0.7969\ \log_{10}(a_0/\mathrm{kpc}) -1.4763\ \log_{10}(\sigma_{\rm hm}/[\mathrm{km\, s^{-1}}])  -2.4074$   & 2 & 0.90 & 0.89 & 0.08 & 1.1 & 0.35 \\
        \relationbf & $\log_{10}(\Delta t_{\rm df}/\mathrm{Gyr}) $ &$=+ 0.8604\ \log_{10}(a_0/\mathrm{kpc}) -0.2999\ \log_{10}(M_{\rm BH,FP}/M_\odot) + 0.9355$ & 2 & 0.84 & 0.82 & 0.10 & 1.6 & 0.45 \\
        \hline
        \multicolumn{9}{c}{\textbf{Hardening}} \\
        \relation & $\log_{10}(\Delta t_{\rm h}/\mathrm{Gyr}) $ &$=+0.9130\ \log_{10}(M_{\rm bin}/M_\odot) + 0.8274\ \log_{10}(r_{\rm peri,h}/\mathrm{kpc})$ & 3 & 0.70 & 0.66 & 0.28 & 1.5 & 0.66 \\
         &  &$\quad \quad \quad \quad \quad \quad \quad \quad \quad \quad \quad \quad \quad \quad -1.6180\ \log_{10}(a_{\rm h}/\mathrm{kpc}) -9.0845$ & & & & & & \\
        \relation & $\log_{10}(\Delta t_{\rm h}/\mathrm{Gyr}) $ &$=+ 2.1145\ \log_{10}(M_{\rm bin}/M_\odot) -3.6317\ e_{\rm b}-1.5032\ \log_{10}(a_{\rm h}/\mathrm{kpc})$ & 4 & 0.81 & 0.78 & 0.22 & 2.4 & 0.51 \\
         &  &$\quad \quad \quad \quad \quad \quad \quad \quad \quad \quad \quad \quad -5.5283\ \log_{10}(\sigma_{\rm 5inf}/[\mathrm{km\, s^{-1}}]) -21.9761$ & & & & & & \\
        \relation & $\log_{10}(\Delta t_{\rm h}/\mathrm{Gyr}) $ &$=-0.4904\ \log_{10}(a_{\rm b}/\mathrm{kpc}) -1.4463\ \log_{10}(\beta_{5\rm inf})-2.7765\ e_{\rm b}$ & 4 & 0.72 & 0.67 & 0.27 & 1.1 & 0.65 \\
         &  &$\quad \quad \quad \quad \quad \quad \quad \quad \quad \quad \quad \quad \quad \quad \quad \quad \quad \quad + 0.7630\ (r_{\rm inf}/\mathrm{kpc})+ 0.8533$ & & & & & & \\
        \relationbf & $\log_{10}(\Delta t_{\rm h}/\mathrm{Gyr}) $ &$=+2.3411\ \log_{10}(M_{\rm bin}/M_\odot) -2.1387\ e_{0}-0.2228\ q_{\star}$ & 4 & 0.45 & 0.35 & 0.38 & 4.3 & 0.87 \\
         &  &$\quad \quad \quad \quad \quad \quad \quad \quad \quad -1.8110\ \log_{10}(M_{\rm enc,rem}(5\, r_{\rm inf})/M_\odot)-0.5027$ & & & & & & \\
        \hline
        \multicolumn{9}{c}{\textbf{Total coalescence time}} \\
        \relation & $\log_{10}(t_{\rm coal}/\mathrm{Gyr}) $ &$= +0.7881\ \log_{10}(a_0/\mathrm{kpc}) -1.4053\ e_{\rm h}$ & 3 & 0.89 & 0.87 & 0.10 & 1.0 & 0.45 \\
         &  &$\quad \quad \quad \quad \quad \quad \quad \quad \quad \quad -1.2615\ \log_{10}(\sigma_{\rm 5inf}/[\mathrm{km\, s^{-1}}]) -0.5522$ & & & & & & \\
        \relationbf & $\log_{10}(t_{\rm coal}/\mathrm{Gyr}) $ &$= +1.0029\ \log_{10}(a_0/\mathrm{kpc}) -0.9977\ e_0$ & 3 & 0.85 & 0.83 & 0.12 & 1.4 & 0.45 \\
         &  &$\quad \quad \quad \quad \quad \quad \quad \quad -0.1756\ \log_{10}(M_{\rm enc,SP}(5\, r_{\rm inf})/M_\odot) + 1.1724$ & & & & & & \\
        \relation & $\log_{10}(t_{\rm coal}/\mathrm{Gyr}) $ &$= +0.5221\ \log_{10}(a_0/\mathrm{kpc}) -1.3153\ e_{\rm h}-0.9378\ \log_{10}(\beta_{5\rm inf})$ & 4 & 0.91 & 0.89 & 0.09 & 1.7 & 0.36 \\
         &  &$\quad \quad \quad \quad \quad \quad \quad \quad -0.1417\ \log_{10}(\rho_{\rm rem}(5\, r_{\rm inf})/[M_\odot\, \mathrm{kpc}^{-3}]) + 1.1514$ & & & & & & 
        \\
        \relation & $\log_{10}(t_{\rm coal}/\mathrm{Gyr}) $ &$= +0.6953\ \log_{10}(a_0/\mathrm{kpc}) -1.4250\ e_{\rm h}-0.5336\ \log_{10}(\beta_{5\rm inf})$ & 4 & 0.92 & 0.90 & 0.09 & 1.3 & 0.34 \\
         &  &$\quad \quad \quad \quad \quad \quad \quad \quad -1.1474\ \log_{10}(\sigma_{5\rm inf}/[\mathrm{km}\, \mathrm{s}^{-1}]) -0.6240$ & & & & & & 
        \\
        \relation & $\log_{10}(t_{\rm coal}/\mathrm{Gyr}) $ &$= +0.7689\ \log_{10}(a_0/\mathrm{kpc}) -1.3671\ e_{\rm h}-0.4570\ \log_{10}(M_{\rm enc,SP}(5\, r_{\rm inf})/M_{\odot})$ & 4 & 0.91 & 0.89 & 0.09 & 2.2 & 0.37 \\
         &  &$\quad \quad \quad \quad \quad \quad \quad \quad +0.4604\ \log_{10}(r_{\rm inf}/\mathrm{kpc}) + 4.9910$ & & & & & & 
        \\
        \relationbf & $\log_{10}(t_{\rm coal}/\mathrm{Gyr}) $ &$= +0.9574\ \log_{10}(a_0/\mathrm{kpc}) -0.8382\ e_0+0.3591\ \log_{10}(r_{\rm inf}/\mathrm{kpc})$ & 4 & 0.88 & 0.86 & 0.10 & 2.8 & 0.37 \\
         &  &$\quad \quad \quad \quad \quad \quad \quad \quad -0.2960\ \log_{10}(M_{\rm enc,SP}(r_{\rm inf})/M_\odot) + 2.1342$ & & & & & & 
        \\
        \relationbf & $\log_{10}(t_{\rm coal}/\mathrm{Gyr}) $ &$= +0.9339\ \log_{10}(a_0/\mathrm{kpc}) -0.8077\ e_0+0.2984\ \log_{10}(r_{\rm inf}/\mathrm{kpc})$ & 4 & 0.86 & 0.84 & 0.11 & 2.8 & 0.37 \\
         &  &$\quad \quad \quad \quad \quad \quad \quad \quad -0.3312\ \log_{10}(M_{\rm enc,SP}(5\, r_{\rm inf})/M_\odot) + 2.8326$ & & & & & & 
        \\

        \hline
    \end{tabular}
\end{table*}

\subsection{Cosmological merger time statistics}
\label{sec:cosmoStat}

\begin{figure}
    \includegraphics[width=\columnwidth]{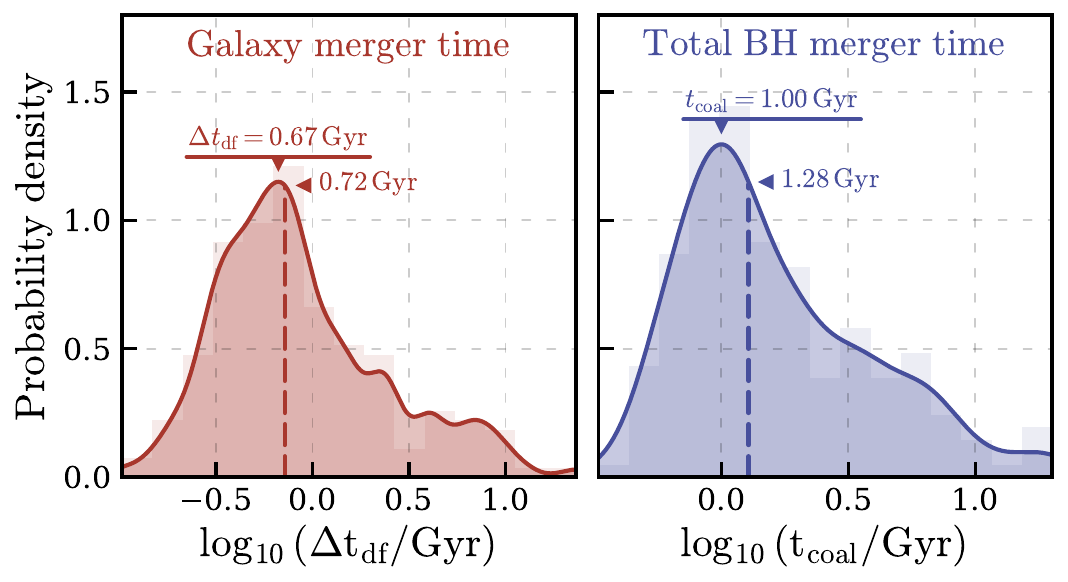}
    \caption{Distributions of predicted galaxy merger times $\Delta t_{\rm df}$ (left) and total BH coalescence times $t_{\rm coal}$ (right) for the 174 mergers selected from \textsc{IllustrisTNG} (TNG50, TNG100, and TNG300) at $z \leq 1$, estimated using relations~(\ref{rel:c}) and (\ref{rel:q}). The distributions are constructed as an equal-weight Gaussian mixture, with one component centred on each individual prediction and a fixed width set by the intrinsic scatter of the relations. The peak value of galaxy merger and BH total coalescence times are $0.67$ Gyr and $1.00$ Gyr respectively, while the medians, shown as dashed lines, are $0.72$ Gyr and $1.28$ Gyr. This indicates that most systems merge within a few Gyr which is consistent with constraints from the GWB.}
    \label{fig:densityTmimesCosmo}
\end{figure}

We now apply relation (\ref{rel:c}) and (\ref{rel:q}) for the dynamical friction time (equivalent to the galaxy merger time) and total SMBH coalescence time directly to the \textsc{IllustrisTNG} simulations using a larger sample of mergers. We select all galaxies at redshift $z = 0$ with stellar mass $M_\star > 10^{11}\, M_\odot$, compared to $3\times10^{11}\, M_\odot$ in the previous selection, in TNG50, TNG100, and TNG300. We then reconstruct their merger histories following the procedure described in Section~\ref{sec:mergerSel}, and we retain mergers occurring at $z \leq 1$ with stellar mass ratios $q_\star > 1/4$. We further exclude unbound encounters, as well as mergers involving galaxies without central SMBHs or systems in which the resulting binary does not satisfy the conditions $q_{\rm BH} > 1/4$ and ${\cal M} > 10^{8}\,M_\odot$. This selection yields a total of 174 mergers.

For each system, we compute the parameters entering relations~(\ref{rel:c}) and (\ref{rel:q}) (see Section~\ref{sec:guide} for details) and use these relations to estimate the galaxy merger time and the SMBH total coalescence time. The resulting distributions, shown in Fig.~\ref{fig:densityTmimesCosmo}, are constructed as an equal-weight Gaussian mixture, with one component centred on each prediction and a fixed width set by the intrinsic scatter of the relations.

We find that the galaxy merger phase, from apocentre to the formation of a SMBHB, typically occurs over $\sim 0.67\, \mathrm{Gyr}$. The binary then hardens and coalesces, with a total time from apocentre to coalescence of $\sim 1\,\mathrm{Gyr}$. The majority of black hole mergers therefore occur on relatively short timescales, consistent with constraints from the GWB, which suggest that SMBHBs typically merge within $\sim 1\, \mathrm{Gyr}$ following galaxy mergers \citep[][]{EPTA2024}. Our results indicate that $\sim 1\, \mathrm{Gyr}$ corresponds to the most probable total coalescence time, i.e. typically $\sim 0.3\, \mathrm{Gyr}$ after the completion of the galaxy merger. The distribution nevertheless exhibits a tail extending beyond this value, corresponding to a population of longer-lived systems.

\section{Conclusions}
\label{sec:conclusion}

\subsection{Summary}

This work aims to characterise the connection between the coalescence time of SMBHBs and the properties of their host galaxies and galaxy encounters, in order to improve the interpretation of GWB observations. The amplitude and spectral shape of this signal are highly sensitive to SMBH coalescence timescales, which are commonly estimated using semi-analytical models that may rely on idealised assumptions and simplified initial conditions, potentially limiting their applicability in a cosmological context.
In this study, we instead adopt a complementary approach by extracting initial conditions for galactic mergers from the \textsc{IllustrisTNG} cosmological simulations and re-simulating them at high resolution using the $N$-body code \textsc{Griffin}. This allows us to accurately capture the dynamical friction phase and the subsequent hardening of the SMBHB. We then couple these simulations to a semi-analytical treatment to account for GW emission in the late stages, enabling us to follow the binary evolution down to coalescence.
Based on this framework, we derive scaling relations linking the duration of the dynamical friction phase, the hardening phase, and the total SMBH coalescence time to the properties of the host galaxies and their orbital configurations. These relations can be readily implemented in subgrid models for SMBHB coalescence times. We then apply them back to the \textsc{IllustrisTNG} simulations, using a larger sample of galaxy mergers, in order to infer the cosmological distribution of SMBH merger timescales.

\noindent The \textbf{main results} can be summarised as follows:
\begin{itemize}
  \item[$\blacksquare$]  \textbf{Initial conditions.} $\quad$ Cosmological merger populations are characterised by highly eccentric galaxy encounters at infall, with a strongly peaked distribution at $e_0 = 0.94$, indicating that near-radial orbits dominate in realistic cosmological settings. Although the sample is selected to include major stellar mergers ($q_\star \gtrsim 1/4$), the corresponding dark matter mass ratios are significantly lower, typically $q_{\rm DM} \sim 0.003$--$0.06$, due to substantial dark matter stripping of the secondary progenitors prior to coalescence due to their cosmological environment. This highlights that idealised mergers obtained by simple rescaling of progenitors do not generally reproduce cosmological conditions.

  \item[$\blacksquare$]  \textbf{Binary evolution and timescales.} $\quad$ SMBHB form with intrinsically high eccentricities, with a peak value $e_{\rm b} = 0.91$. The eccentricity further increases during the subsequent hardening phase, leading to the emergence of a more pronounced high-eccentricity excess in the $e_{\rm h}$ distribution at the hard-binary separation. This underscores the importance of non-circular configurations throughout the evolution. In terms of timescales, binaries typically spend about twice as long in the dynamical friction phase as in the stellar hardening phase, and about twice as long in the hardening phase as in the GW phase, implying that dynamical friction dominates the overall evolution. The total coalescence time of the simulated binaries is typically $\sim 1\,\mathrm{Gyr}$.

  \item[$\blacksquare$]  \textbf{PTA detection.} $\quad$ Binaries enter the PTA band with extremely high eccentricities, strongly peaked at $e_{\rm PTA} = 0.992$, and remain in the band for a typical residence time of $\sim 131\,\mathrm{Myr}$.

  \item[$\blacksquare$]  \textbf{Relations.} $\quad$ We derive empirical relations linking SMBHB evolution timescales to host galaxy and orbital properties. A comprehensive guide to the implementation and recommended usage of these relations is provided in Section~\ref{sec:guide}. The dynamical friction phase is primarily controlled by the semi-major axis of the galaxy encounter and, when combined with the central velocity dispersion, yields nearly perfect predictive power ($R^2 = 0.99$). An alternative, simpler relation with the primary black hole mass is also provided, achieving $R^2 = 0.94$. The hardening phase depends on two binary orbital parameters, the binary mass, and one central property of the remnant (density or velocity dispersion), with predictive power ranging from $R^2 = 0.70$ to $0.81$, and degrading when these quantities are not accessible ($R^2 = 0.45$). The total coalescence time can be robustly predicted, with $R^2 = 0.89$--$0.92$, and remains accurate even when restricted to quantities available in cosmological simulations ($R^2 = 0.85$--$0.88$). These relations provide practical subgrid prescriptions for SMBH merger timescales based on the properties of their progenitor galaxies.
  
  \item[$\blacksquare$]  \textbf{Cosmological statistics.} $\quad$ Applying these relations to a larger sample of mergers from \textsc{IllustrisTNG} shows that galaxy mergers typically complete within $\sim 0.7\,\mathrm{Gyr}$, while SMBHBs coalesce over a total timescale of $\sim 1\,\mathrm{Gyr}$. These results are in excellent agreement with constraints from PTAs, which suggest that massive black hole mergers generally take place within $\sim 1\,\mathrm{Gyr}$ following the merger of their host galaxies.
\end{itemize}

\subsection{Comparison with previous work and discussion}

The eccentricity of our binaries at the point of entry into the PTA frequency band 
(Fig.~\ref{fig:histoPTA}) can be compared with the findings of \citet{Fastidio2025}. Our binaries exhibit significantly higher eccentricities at this stage, with a peak value of $e_{\rm PTA} = 0.992$, which exceeds the eccentricity of the most eccentric binary in \citet{Fastidio2025}. This discrepancy may in part reflect the small sample size of that study, which comprises only eight binaries, and may therefore not be representative of the full population. We note, however, that both studies consistently find highly eccentric orbits at PTA entry, with all systems in \citet{Fastidio2025} having $e_{\rm PTA} > 0.86$, suggesting that high 
eccentricity at PTA entry may be a robust feature of massive black hole binary evolution, regardless of the exact sample.

We next compare our scaling relations with those derived by \citet{HolleyBockelmann2025}. In particular, we attempt to recover their four-parameter relation within our framework. In their study, the simulations are initialised with SMBHs separated by a few influence radii, and the relations are constructed between this initial configuration and the final coalescence time. To facilitate a meaningful comparison, we consider the time interval between binary formation and coalescence, $[t_{\rm b}, t_{\rm coal}]$, where $t_{\rm b}$ denotes the time at which the binary becomes bound and $t_{\rm coal}$ the time of merger. We then search for a relation between this duration and a set of parameters analogous to those entering their model. The closest relation we obtain is given by:
\begin{equation}
\begin{split}
    \log_{10}(t_{\rm coal}/\mathrm{Gyr}) = & \quad \, 2.0676\, \log_{10}(M_{\rm bin}/M_\odot) \\&-1.5656\, \log_{10}(M_{\rm enc,rem}(5\, r_{\rm inf})/M_\odot) \\&+ 0.8176\, (\left . \bar{v}_{\varphi}/\sigma_{\varphi}\right |_{\rm hm}) -0.0061\, r_{\rm hm}-2.7890
\end{split}
\end{equation}
which  involves parameters that are analogous to those of \citet{HolleyBockelmann2025}. However, in our setup, its predictive power is found to be very limited, with a coefficient of determination of $R^2 = 0.13$.

We therefore explored more general forms by introducing additional degrees of freedom in the parameter choice while attempting to preserve the underlying physical content of the original model. In particular, we considered combinations of parameters describing (i) the central mass structure of one of the progenitors or of the remnant, (ii) the mass of the SMBHs or of the binary, (iii) the stellar kinematics, and (iv) a characteristic size of the system. Even within this extended parameter space, the best-fitting relation we obtain is given by:
\begin{equation}
\begin{split}
    \log_{10}(t_{\rm coal}/\mathrm{Gyr}) = &\quad \,2.3789\, \log_{10}(M_{\rm BH,FP}/M_\odot) \\&-0.5906\, \log_{10}(M_{\rm enc,NP}( r_{\rm inf})/M_\odot) \\&-4.6882\, \log_{10}(\sigma_{\rm inf}/[\mathrm{km}\, \mathrm{s}^{-1}]) -0.0249\, r_{\rm hm}\\&-17.8745.
\end{split}
\end{equation}
This best relation yields a coefficient of determination of $R^2 = 0.24$, indicating only a modest improvement in predictive power.

We note, however, that allowing for the inclusion of a single additional parameter, namely the binary eccentricity at formation, $e_{\rm b}$, leads to a significant enhancement. In this case, we obtain the following relation:
\begin{equation}
\begin{split}
    \log_{10}(t_{\rm coal}/\mathrm{Gyr}) = &\quad \,2.3375\, \log_{10}(M_{\rm BH,FP}/M_\odot) -3.8832\, e_{\rm b} \\&-6.1546\, \log_{10}(\sigma_{\rm 5inf}/[\mathrm{km}\, \mathrm{s}^{-1}]) -0.0406\, r_{\rm hm}\\&-20.0642
\end{split}
\end{equation}
which reaches a coefficient of determination of $R^2 = 0.48$.

The fact that \citet{HolleyBockelmann2025} are able to obtain high predictive power without including an explicit dependence on the binary eccentricity at formation, $e_{\rm b}$, can be understood in light of their specific numerical setup. In their study, the host galaxy is endowed with significant rotational support and the same merger configuration is explored under different orbital senses, with the secondary black hole placed on either prograde or retrograde orbits (all with fixed eccentricity $e = 0.5$). This choice leads to a pronounced bimodality in the eccentricity of the binary at formation: prograde configurations systematically produce nearly circular binaries, with $e_{\rm b} \sim 0$--$0.2$, whereas retrograde configurations yield highly eccentric binaries, with $e_{\rm b} \sim 0.8$--$1$. Residual variations within a given configuration are likely driven by stochastic effects \citep[see][]{Rawlings2023, Gualandris2026}. In this context, the sense of rotation effectively encodes the relevant orbital information of the binary, which is a key parameter in determining the hardening timescale. This phase, in turn, dominates the overall evolution time during the late stages of the inspiral.

More generally, in systems with significant rotational support, rotation may therefore act as a proxy for the binary orbital configuration, alleviating the need for an explicit dependence on $e_{\rm b}$ (and $a_{\rm b}$). However, this interpretation may be less applicable to the population of PTA sources, which are expected to reside predominantly in massive elliptical galaxies with low levels of ordered rotation. In such systems, rotation is unlikely to provide a strong constraint on the binary orbital properties, as we saw in this paper, and may therefore play a limited role in shaping the GWB.

We nevertheless caution against over-interpreting the apparent predictive power of rotation in simplified setups. Two key assumptions remain to be tested in more realistic merger configurations. First, it is unclear whether the coupling between galaxy rotation and binary orbital properties remains as efficient for the highly radial encounters that characterise cosmological mergers. In this regime, additional processes may become important. For instance, \citet{Boily2025} highlight the role of dynamical traction in transferring angular momentum to initially radial black hole orbits, potentially leading to orbital circularisation and delayed coalescence. Second, the setup of \citet{HolleyBockelmann2025} considers the inspiral of a single black hole into a pre-existing galaxy, whereas in realistic mergers the secondary black hole is embedded within the stellar nucleus of its progenitor. The presence of this bound structure may alter the interaction with the host galaxy and could potentially reduce the strong bimodality in $e_{\rm b}$ reported in their study.

Finally, regardless of the rotation, a residual level of stochasticity appears to be unavoidable in the binary evolution, whether physical in origin \citep[][]{Rawlings2023} or arising from resolution effects \citep[][]{Gualandris2026}. While such stochastic effects inevitably limit the precision of predictions, meaningful constraints on SMBHB coalescence timescales can nonetheless be obtained. This is primarily because the dynamical friction phase contributes significantly to the total evolution time and can be robustly linked to large-scale galaxy properties. Although the hardening phase exhibits stochasticity, it retains a non-negligible degree of predictability through its dependence on progenitor and orbital properties prior to the merger (e.g. the role of $e_0$ in relation~\ref{rel:j}). Taken together, these effects are particularly encouraging for PTA studies, as they lead to relations that perform well in predicting SMBHB coalescence timescales, even in the presence of stochasticity and limited constraints on the binary orbital configuration.

\subsection{Caveats}

Several limitations of our approach should be kept in mind when interpreting the results presented in this work.

We do not include gas dynamics in our simulations, and our results therefore primarily apply to gas-poor massive galaxies, which constitute the dominant population of PTA sources. However, a fraction of such systems may still contain non-negligible gas, in which case additional physical processes could alter the binary evolution and would require dedicated modelling.

In addition, our modelling framework relies on logarithmic transformations of the variables, which enables us to capture power-law dependencies in the scaling relations. However, this approach does not account for more complex functional forms, such as exponential dependencies or non-separable relations between variables, including higher-order interaction terms. Given the relatively small size of the sample (30 simulations), the statistical power is insufficient to robustly constrain more complex models, and the risk of overfitting would be significant. Exploring such functional forms would require either a substantially larger dataset or strong a priori physical motivation for a specific parametrization, which remains challenging in this context. Despite its simplicity, our approach offers good interpretability and robustness, and already provides satisfactory predictions of the merger timescales.

Although we restrict the analysis to low-dimensional models (one to four parameters) and apply stringent selection criteria based on $R^2$, variance inflation factors (VIF), and cross-validation, the large number of models explored relative to the sample size implies that some of the identified correlations may be driven by statistical fluctuations (the so-called look-elsewhere effect). The resulting scaling relations should therefore be regarded as indicative rather than definitive, particularly for those with moderate values of $R^2$, while the most predictive relations are expected to be more robust.

The present analysis is restricted to major mergers ($q_\star \gtrsim 1/4$), which are expected to dominate the contribution to the GWB. Minor mergers ($q_\star < 1/4$) are, however, significantly more numerous and may contribute non-negligibly to the overall merger rate. Extending our framework to this regime is not straightforward. While the orbital parameters at infall ($a_0,e_0$) are primarily set by cosmological statistics and are not expected to depend strongly on the mass ratio, the dynamical friction phase may behave qualitatively differently at low $q_\star$. In particular, the secondary galaxy is more susceptible to tidal disruption before its nucleus reaches the centre of the primary, potentially causing the secondary SMBH to decouple from its stellar host \citep[][]{bellovary2010}, a process not captured by our scaling relations. The hardening phase is also likely to be affected: minor mergers induce less violent relaxation and therefore preserve the central density more effectively, which may accelerate hardening and is in principle partially captured by the dependence of our relations on central density and velocity dispersion. However, minor mergers are more likely to involve gas-rich second progenitors that can modify the black hole mass ratio \citep[][]{callegari2011}, dynamical friction and hardening evolution \citep[][]{escala2005, dotti2007} in ways not accounted for by our purely $N$-body framework. Our scaling relations may therefore not be valid in the case of minor mergers.

\subsection{Guide to using the scaling relations}
\label{sec:guide}

This section is intended to guide readers who wish to apply the provided scaling relations to cosmological simulations. We discuss both the choice of the appropriate relations and the practical computation of the relevant parameters, ensuring consistency with the definitions and quantities adopted in this work.

\subsubsection{You want to estimate the galaxy merger (dynamical friction) time}

\noindent \textbf{Remnant galaxy available.} 
Use relation~(\ref{rel:b}). This relation depends on the semi-major axis of the galaxy encounter, $a_0$, derived from equation~(\ref{eq:galacticOrbit}), and on the one-dimensional stellar velocity dispersion of the remnant within the half-mass radius of the primary progenitor, $\sigma_{\rm hm}$. The relative velocity $v_{\rm rel}$ entering the definition of $a_0$ is evaluated after subtracting the bulk motions measured within the stellar half-mass radii of the progenitors.

\noindent \textbf{Only progenitors available.} 
Use relation~(\ref{rel:c}), which replaces $\sigma_{\rm hm}$ with the black hole mass of the primary progenitor, $M_{\rm BH,FP}$, while still requiring $a_0$.

\subsubsection{You want to estimate the binary hardening time}

\noindent \textbf{Binary and remnant available.} 
Use relation~(\ref{rel:i}). This relation depends on the binary semi-major axis and eccentricity at formation, $a_b$ and $e_b$. Note that shortly after binary formation the eccentricity may undergo transient oscillations before settling to a stable value. We therefore define $e_b$ as the eccentricity measured at its first local minimum following this oscillation phase, and $a_b$ is taken at the same time. 
It also requires the influence radius $r_{\rm inf}$, defined such that the enclosed stellar mass in the primary equals the mass of the more massive black hole, and the anisotropy parameter $\beta_{5\rm inf}$ measured within $5\,r_{\rm inf}$. To compute $\beta_{5\rm inf}$, the $z$-axis must be aligned with the total angular momentum of stars enclosed within this radius.

\noindent \textbf{Binary not available.} 
Relation~(\ref{rel:j}) can be used as an approximate estimate. It depends on the binary mass $M_{\rm bin}$, the orbital eccentricity of the galactic encounter $e_0$ (see equation~\ref{eq:galacticOrbit}), the stellar mass ratio $q_{\star}$, and the enclosed mass of the remnant within $5\,r_{\rm inf}$. This relation should be used with caution as without access to the binary parameters deriving robust relations proves difficult. This relation may therefore serve as a rough indicator, but we caution against relying on it when precise and robust predictions are required.

\subsubsection{You want to estimate the total coalescence time}

We provide three relations that depend solely on the properties of the progenitor galaxies (relations~\ref{rel:l}, \ref{rel:p}, and~\ref{rel:q}). All require the orbital parameters of the galactic encounter $a_0$ and $e_0$, computed from equations~(\ref{eq:galacticOrbit}). Relation~(\ref{rel:l}) depends on three parameters, including the enclosed mass of the secondary progenitor within $5\,r_{\rm inf}$, $M_{\rm enc,SP}(5\,r_{\rm inf})$. Relation~(\ref{rel:p}) and (\ref{rel:q}) add $r_{\rm inf}$ as an additional parameter, the only difference between the two being that relation~(\ref{rel:q}) keeps $M_{\rm enc,SP}(5\,r_{\rm inf})$ while relation~(\ref{rel:p}) instead uses the enclosed mass at $r_{\rm inf}$, $M_{\rm enc,SP}(r_{\rm inf})$. We recommend relation~(\ref{rel:p}) when the inner density profile of the secondary progenitor is well resolved. Otherwise, relation~(\ref{rel:l}) and (\ref{rel:q}) provide a more robust alternative with the latter yielding a marginal improvement in performance at the cost of one additional parameter.

\subsubsection{Practical computation of input quantities}

To compute the enclosed masses of the galaxies, we fit the density profiles from the cosmological simulation snapshot using a MCMC approach. The adopted profile is a truncated spheroid of the form given by equation~(\ref{eq:denProfile}). The enclosed masses, the half-mass radius, and the influence radius are then computed from these fitted density profiles rather than directly from particle data. This approach effectively performs a spherical average of the galaxy.

\section*{Acknowledgements}
AG acknowledges support from grant ST/Y002385/1. The simulations were run on the Eureka2 HPC cluster at the University of Surrey.
This research made use of the following software: \textsc{Agama} \citep[][]{Vasiliev2019}, \textsc{Emcee} \citep[][]{emcee}, \textsc{Griffin} \citep[][]{Dehnen2014}, \textsc{Jupyter} \citep{jupyter}, \textsc{Matplotlib} \citep{Matplotlib}, \textsc{Numpy} \citep{Numpy}, \textsc{Pandas} \citep[][]{pandas}, \textsc{Python} \citep[][]{python}, \textsc{Scikit-learn} \citep[][]{ScikitLearn} and \textsc{Scipy} \citep[][]{scipy}.

\section*{Data Availability}

The authors will share the data underlying this article upon reasonable request.



\bibliographystyle{mnras}
\bibliography{example} 




\appendix

\section{Number of particles and resolution}
\label{sec:Resolution}

\begin{figure*}
	\centering
	\includegraphics[width=\textwidth]{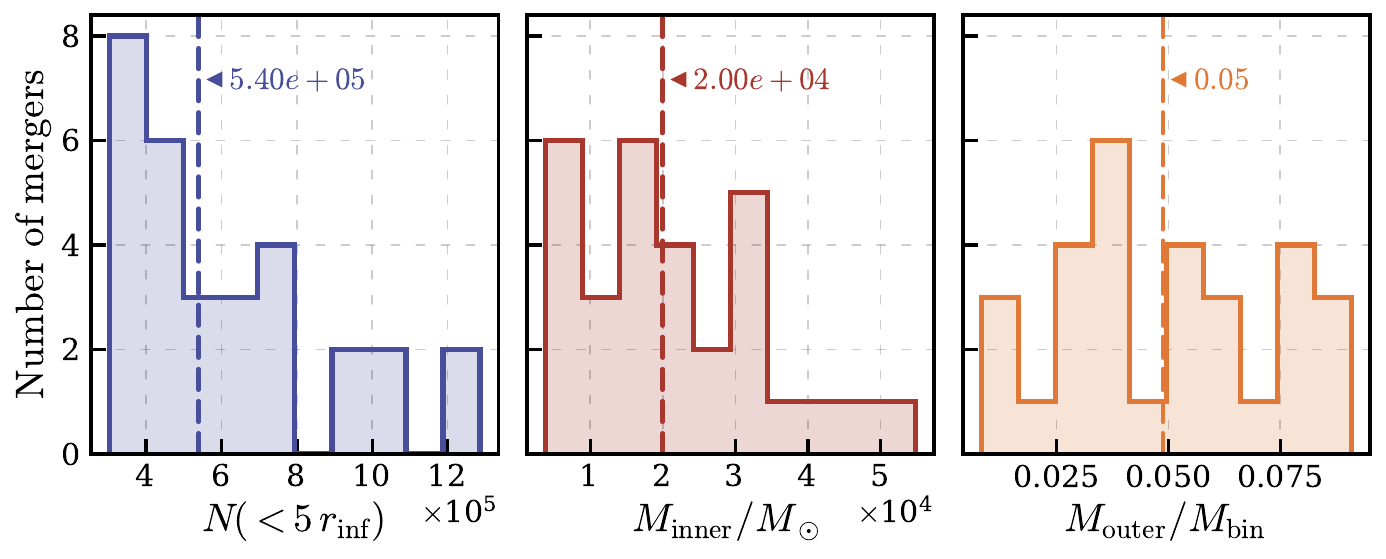}
    \caption{Resolution metrics for the 30 simulations analysed in this work. The left panel shows the distribution of the number of first-progenitor particles enclosed within five times the binary radius of influence $r_\mathrm{inf}$, at $t = 0$ Gyr for each simulation. The middle panel shows the particle mass of the most central and best-resolved population in each run, $M_\mathrm{inner}$. The right panel shows the mass of the most massive particles, $M_\mathrm{outer}$, located in the outer regions, normalised by the mass of the black hole binary $M_\mathrm{bin}$. The adopted mass refinement scheme yields very high resolution in the central regions while keeping the masses of outer particles acceptable. In all panels, median values are indicated by dashed lines, with the corresponding values labelled.}
    \label{fig:Resolution}
\end{figure*}

Figure~\ref{fig:Resolution} summarises the resolution choices adopted for the full set of 30 simulations presented in this work.
The left-hand panel shows the number of particles belonging to the first progenitor enclosed within five times the binary influence radius at $t=0$~Gyr. Across the full sample, this number typically ranges between $\sim 4\times10^{5}$ and $\sim 12\times10^{5}$ particles. Higher particle numbers were preferentially adopted for highly eccentric encounters, which experience very small pericentric passages. In such cases, insufficient resolution has been shown to introduce stochastic fluctuations in the binary eccentricity at formation ($e_\mathrm{b}$) \citep[][]{Gualandris2026}. We note, however, that the resolution required to achieve a certain level of convergence in $e_\mathrm{b}$ depends non-linearly on several parametres, including densities and mass ratio. Given the diversity of merger properties in our sample, defining a unique, case-by-case convergence criterion for the eccentricity across all 30 simulations remains challenging.

The central panel displays the mass of the best-resolved particles in the innermost region. In most simulations, particle masses reach values of a few $10^{4}\, M_\odot$, ensuring high resolution in the regions most relevant for the formation and evolution of the SMBH binary. Increasing the mass resolution in the central regions necessarily comes at the expense of larger particle masses in the outer dark matter shells. Care is therefore taken to limit the mass of these outer particles in order to avoid numerical artefacts. 

The right-hand panel shows the mass of the most massive particles relative to the total mass of the SMBHB. In all simulations, these particles remain sufficiently light compared to the binary to prevent spurious perturbations of its orbital evolution.

\section{Black hole to total stellar mass relation}
\label{sec:BHstarRelation}

\begin{figure}
    \includegraphics[width=\columnwidth]{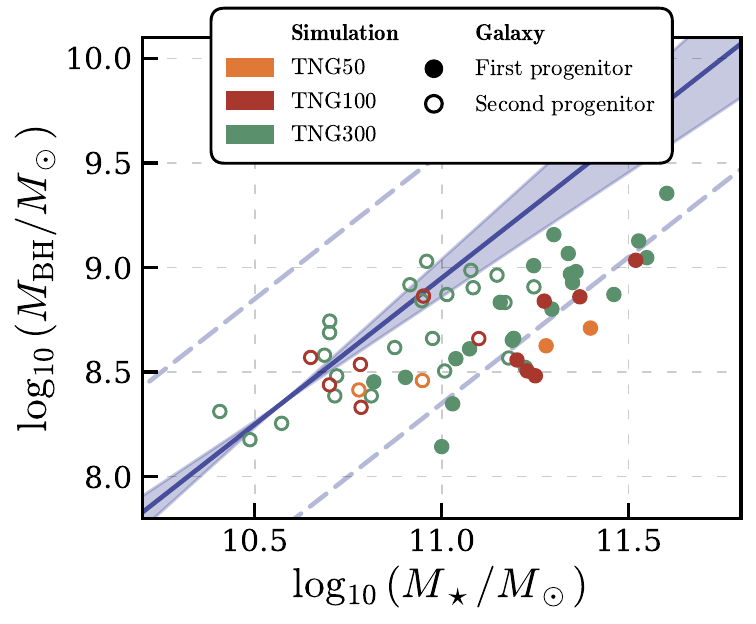}
    \caption{Black hole mass as a function of total stellar mass for the selected \textsc{IllustrisTNG} merger sample. Galaxies from the \textsc{IllustrisTNG50}, \textsc{TNG100}, and \textsc{TNG300} volumes are shown in orange, red, and green, respectively. Filled markers correspond to first progenitors, while open markers indicate second progenitors. The empirical black hole–stellar mass relation from \citet{Reines2015} is overplotted for comparison. Overall, the black holes in our sample lie slightly below the observed relation.}
    \label{fig:MbhRelationTNG}
\end{figure}

Figure~\ref{fig:MbhRelationTNG} shows the $(M_{\rm BH}, M_\star)$ pairs for our selected sample of \textsc{IllustrisTNG} mergers. For comparison, we also plot the black hole–to–total stellar mass relation reported by \citet{Reines2015} for elliptical galaxies,
\begin{equation}
    \log\left(\frac{M_{\text{BH}}}{M_\odot}\right) = \alpha + \beta\ \log\left(\frac{M_\star}{10^{11} M_\odot}\right)
\end{equation}
with best-fitting parameters
\begin{equation}
    \alpha = 8.95 \pm 0.09\quad ; \quad \beta = 1.40 \pm 0.21.
\end{equation}

\section{Total coalescence times distribution}
\label{sec:densityTmerg}

\begin{figure}
    \includegraphics[width=\columnwidth]{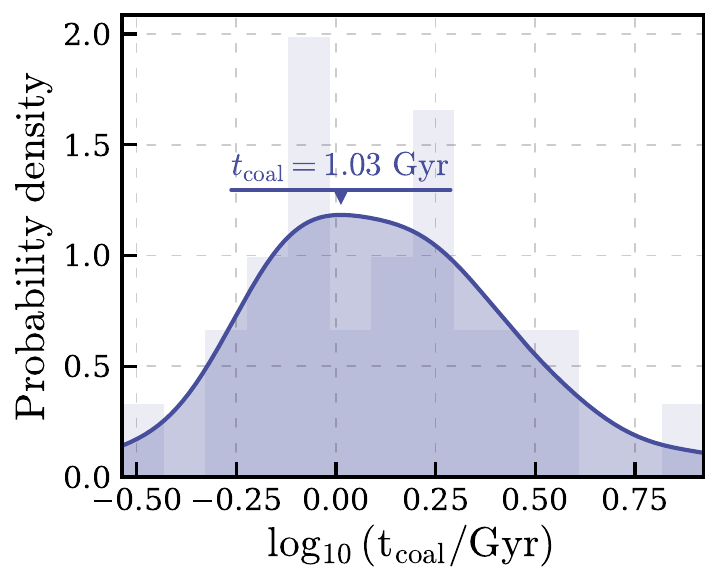}
    \caption{Probability density of the coalescence times $t_{\rm coal}$ of our simulated binaries, estimated using a kernel density estimator. The distribution peaks at $t_{\rm coal}=1.03$~Gyr.}
    \label{fig:densityTmerg}
\end{figure}

Figure~\ref{fig:densityTmerg} shows the probability density of coalescence times 
of our simulated binaries, estimated using a kernel density estimator. The distribution peaks at $t_{\rm coal} = 1.03$ Gyr, indicating that most binaries coalesce on relatively short timescales.


\bsp	
\label{lastpage}
\end{document}